\documentclass[pra,aps,superscriptaddress,twocolumn]{revtex4-2}
\usepackage{graphicx} 
\usepackage{amsmath,amssymb, theorem} 
\usepackage{physics}
\usepackage{tikz}
\usepackage{tikz-network}
\usepackage{xifthen}
\usepackage{xargs}

\newcommandx{\fineq}[5][1=-.8ex,2=1,3=1,5=0]{\begin{tikzpicture}[baseline={([yshift=#1]current  bounding  box.center)}, scale = #2, every node/.style={scale = #3},rotate around={#5:(0,0)},every node/.style={transform shape}]
		#4
	\end{tikzpicture}
}

\tikzset{
    cross/.pic = {
    \draw[very thick, rotate = 45] (-#1,0) -- (#1,0);
    \draw[very thick, rotate = 45] (0,-#1) -- (0, #1);
    }
}

\definecolor{bertinired}{RGB}{232,102,102}
\definecolor{bertiniblue}{RGB}{101,147,245}
\definecolor{bertinigreyblue}{RGB}{166,218,149}
\definecolor{bertinigreyred}{RGB}{232,102,102}
\definecolor{bertinivioletc}{RGB}{45,130,60}
\definecolor{bertinigreen}{RGB}{166,218,149}
\definecolor{bertiniorange}{RGB}{255, 116, 23}
\definecolor{OliveGreen}{RGB}{85,107,47}
\definecolor{NavyBlue}{RGB}{0,0,128}
\definecolor{bertiniviolet}{RGB}{210,145,178}
\definecolor{bertinigrey1}{RGB}{98,98,98}
\definecolor{bertinigrey2}{RGB}{211,211,211}
\definecolor{bertinigrey3}{RGB}{192,192,192}
\definecolor{bertinigrey4}{RGB}{169,169,169}

\newcommandx{\tikzdiagup}{
	\tikz {\draw[thick] (0,0)--(0.15,0.15); \draw (0,0) rectangle (0.15,0.15);}
}

\newcommandx{\gatecross}[1][1=0.5]{
	\pgfmathparse{#1/2.0}
	\let\x\pgfmathresult
	\draw[thick] (-\x,-\x) -- (\x,\x);
	\draw[thick] (\x,-\x) -- (-\x,\x);
}
\newcommandx{\gatesqu}[2][1=0.25,2=]{
	\pgfmathparse{#1/2.0}
	\let\x\pgfmathresult
	\ifthenelse{\equal{#2}{}}{
		\draw[thick, fill=white, rounded corners=2pt] (-\x,\x) rectangle (\x,-\x);
	}{
		\draw[thick, fill=#2, rounded corners=2pt] (-\x,\x) rectangle (\x,-\x);
	}
}

\newcommandx{\gatemark}[2][1=0.075,2=tr]{
	\pgfmathparse{#1}
	\let\l\pgfmathresult
	\ifthenelse{\equal{#2}{topleft}}{
		\draw[thick] (0,\l) -- ++(-\l,0) --++ (0,-\l);
	}{}
	\ifthenelse{\equal{#2}{topright}}{
		\draw[thick] (0,\l) -- ++(\l,0) --++ (0,-\l);
	}{}
	
	\ifthenelse{\equal{#2}{bottomleft}}{
		\draw[thick] (0,-\l) -- ++(-\l,0) --++ (0,\l);
	}{}
	\ifthenelse{\equal{#2}{bottomright}}{
		\draw[thick] (0,-\l) -- ++(\l,0) --++ (0,\l);
	}{}
	
}

\newcommandx{\roundgate}[5][1=0,2=0,3=1,4=topright,5=white]{
	\pgfmathparse{#3}
	\let\l\pgfmathresult
	\begin{scope}[shift={(#1,#2)}]
		\gatecross[\l]
		
		\pgfmathparse{\l/2.0}
		\let\s\pgfmathresult
		\gatesqu[\s][#5]
		
		\pgfmathparse{\l*0.15}
		\let\m\pgfmathresult
		\gatemark[\m][#4]
	\end{scope}
}
\newcommandx{\wcirc}[2]{\begin{scope}
		\draw[fill=white] (#1,#2) circle (0.15);	\end{scope}} 
\newcommandx{\wcircc}[2]{\begin{scope}
		\draw[fill=white] (#1,#2) circle (0.13);	\end{scope}} 

\newcommandx{\wsqr}[2]{\begin{scope}
		\draw[fill=white,shift={(#1,#2)}] (-.13,.13) rectangle (.13,-.13);	\end{scope}} 
\newcommandx{\wsqrr}[2]{\begin{scope}
		\draw[fill=white,shift={(#1,#2)}] (-.11,.11) rectangle (.11,-.11);	\end{scope}} 
\newcommandx{\bcirc}[2]{\begin{scope}
		\draw[fill=black] (#1,#2) circle (0.15);	\end{scope}} 
\newcommandx{\thetastate}[4][1=0,2=0,3=1,4=]{
	\pgfmathparse{#3/2}
	\let\l\pgfmathresult
	\pgfmathparse{\l*0.15}
	\let\m\pgfmathresult
	\begin{scope}[shift={(#1,#2)}]
		\draw[thick] (0,0)--(\l,\l);
		\draw[thick] (0,0)--(-\l,\l);
		\ifthenelse{\equal{#4}{}}{
			\draw[fill=white] (0,0) circle (0.15);
		}{
			\draw[thick, fill=#4] (0,0) circle (0.15);
		}
	\end{scope}
}

\newcommandx{\thetastateflipped}[4][1=0,2=0,3=1,4=]{
	\pgfmathparse{#3/2}
	\let\l\pgfmathresult
	\pgfmathparse{\l*0.15}
	\let\m\pgfmathresult
	\begin{scope}[shift={(#1,#2)}]
		\draw[thick] (0,0)--(\l,-\l);
		\draw[thick] (0,0)--(-\l,-\l);
		\ifthenelse{\equal{#4}{}}{
			\draw[fill=white] (0,0) circle (0.15);
		}{
			\draw[thick, fill=#4] (0,0) circle (0.15);
		}
	\end{scope}
}

\newcommandx{\vertgate}[5][1=0,2=0,3=4,4=bertiniorange,5=topright]
{
	\begin{scope}[shift={(#1,#2)}]
		\ifthenelse{\equal{#3}{1}}{
			\roundgate[0][0][1][#5][#4]
		}{
			\foreach \n[evaluate=\n as \y using {2*\n-2}] in {1,...,#3}{
				\roundgate[0][\y][1][#5][#4]
			}
		}
	\end{scope}
}

\newcommandx{\tsfmatV}[8][1=0,2=0,3=l,4=4,5=tr,6=init,7=bertiniorange,8=topright]{
	\begin{scope}[shift={(#1,#2)}]
		\ifthenelse{\equal{#3}{l}}{
			\pgfmathsetmacro{\flag}{0}
		}{
			\pgfmathsetmacro{\flag}{1}
		}
		
		\foreach \y[evaluate=\y as \x using {mod(\y+\flag,2)}] in {1,...,#4}{
			\roundgate[\x][\y][1][#8][#7]
		}
		\ifthenelse{\equal{#5}{tr}}{
			\foreach \y[evaluate=\y as \x using {mod(\y+\flag,2)}] in {#4}{
				\draw [fill=white] (\x-0.5,\y+0.5) circle (0.15);
				\draw [fill=white] (\x+0.5,\y+0.5) circle (0.15);
			}
		}{}
		\ifthenelse{\equal{#6}{init}}{
			\thetastate[\flag][0][1][#7]
		}{}
	\end{scope}
}

\newcommandx{\leftriangle}[5][1=0,2=0,3=4,4=bertiniorange,5=topright]{
	\begin{scope}[shift={(#1,#2)}]
		\pgfmathsetmacro{\t}{#3}
		\pgfmathsetmacro{\steps}{ceil(\t/2)}
		\foreach \i[evaluate=\i as \x using -\t+2*\i-1, evaluate=\i as \ylim using \t-2*\i+2] in {1,...,\steps}{
			\foreach \y[evaluate=\y as \thisx using {\x+\y-1}] in {1,...,\ylim}{
				\roundgate[\thisx][\y][1][#5][#4]
			}
		}
	\end{scope}
}

\newcommandx{\rightriangle}[5][1=0,2=0,3=4,4=bertiniorange,5=topright]{
	\begin{scope}[shift={(#1,#2)}]
		\pgfmathsetmacro{\t}{#3}
		\pgfmathsetmacro{\steps}{ceil(\t/2)}
		\foreach \i[evaluate=\i as \x using -\t+2*\i-1, evaluate=\i as \ylim using \t-2*\i+2] in {1,...,\steps}{
			\foreach \y[evaluate=\y as \thisx using {-\x-\y+1}] in {1,...,\ylim}{
				\roundgate[\thisx][\y][1][#5][#4]
			}
		}
	\end{scope}
}

\newcommandx{\eigenVL}[8][1=0,2=0,3=l,4=5,5=tr,6=init,7=bertiniorange,8=topright]{
	\begin{scope}[shift={(#1,#2)}]
		\pgfmathsetmacro{\t}{#4}
		\leftriangle[0][0][\t][#7][#8]
		
		\ifthenelse{\equal{#6}{init}}{
			\drawinitstate[0][0][l][\t][#7]
		}{}
		
		\ifthenelse{\equal{#5}{tr}}{
			\draw[fill=white] \foreach \x in {0,...,\t} {(\x-0.5-\t,0.5+\x) circle (0.15)};
			\ifthenelse{\equal{#3}{r}}{
				\draw[fill=white] (0.5,\t+0.5) circle (0.15);
			}{}
		}{}
		\ifthenelse{\equal{#5}{parttr}}{
			\draw[fill=white] \foreach \x in {0,...,\t} {(\x-0.5-\t,0.5+\x) circle (0.15)};
		}{}
	\end{scope}
}

\newcommandx{\eigenVR}[8][1=0,2=0,3=l,4=5,5=tr,6=init,7=bertiniorange,8=topright]{
	\begin{scope}[shift={(#1,#2)}]
		\pgfmathsetmacro{\t}{#4}
		\rightriangle[0][0][\t][#7][#8]
		
		\ifthenelse{\equal{#6}{init}}{
			\drawinitstate[0][0][r][\t][#7]
		}{}
		
		\ifthenelse{\equal{#5}{tr}}{
			\draw[fill=white] \foreach \x in {0,...,\t}{(-\x+0.5+\t,0.5+\x) circle (0.15)};
			\ifthenelse{\equal{#3}{l}}{
				\draw[fill=white] (-0.5,\t+0.5) circle (0.15);
			}{}
		}{}
	\end{scope}
}

\newcommandx{\tra}[2][1]{\underset{#1}{\text{tr}}\left[#2\right]}

\newcommandx{\tsfmatDgate}[7][1=0,2=0,3=l,4=4,5=tr,6=bertiniorange,7=topright]
{
	\begin{scope}[shift={(#1,#2)}]
		\ifthenelse{\equal{#3}{l}}{
			\pgfmathsetmacro{\flag}{-1}
		}{
			\pgfmathsetmacro{\flag}{1}
		}
		\pgfmathsetmacro{\t}{#4}
		\foreach \i[evaluate=\i as \x using {\flag*\i}, evaluate=\i as \y using \i] in {1,...,\t}{
			\roundgate[\x][\y][1][#7][#6]
		}
		
		\ifthenelse{\equal{#5}{tr}}{
			\foreach \i[evaluate=\i as \x using {\flag*\i}, evaluate=\i as \y using \i] in {\t}{
				\draw [fill=white] (\x-0.5,\y+0.5) circle (0.15);
				\draw [fill=white] (\x+0.5,\y+0.5) circle (0.15);
			}  
		}{}
	\end{scope}
	
}

\newcommandx{\tsfmatD}[8][1=0,2=0,3=l,4=4,5=tr,6=init,7=bertiniorange,8=topright]{
	\begin{scope}[shift={(#1,#2)}]
		\ifthenelse{\equal{#6}{init}}{
			\thetastate[0][0][1][#7]
		}{}
		
		\ifthenelse{\equal{#3}{l}}{
			\pgfmathsetmacro{\flag}{-1}
		}{
			\pgfmathsetmacro{\flag}{1}
		}
		
		\pgfmathsetmacro{\t}{#4}
		\foreach \i[evaluate=\i as \x using {\flag*\i}, evaluate=\i as \y using \i] in {1,...,\t}{
			\roundgate[\x][\y][1][#8][#7]
		}
		
		\ifthenelse{\equal{#5}{tr}}{
			\foreach \i[evaluate=\i as \x using {\flag*\i}, evaluate=\i as \y using \i] in {\t}{
				\draw [fill=white] (\x-0.5,\y+0.5) circle (0.15);
				\draw [fill=white] (\x+0.5,\y+0.5) circle (0.15);
			}  
		}
		\ifthenelse{\equal{#5}{parttr}}{
			\foreach \i[evaluate=\i as \x using {\flag*\i}, evaluate=\i as \y using \i] in {\t}{
				\draw [fill=white] (\x+0.5*\flag,\y+0.5) circle (0.15);
			}  
		}
		{}
	\end{scope}
}

\newcommandx{\drawinitstate}[5][1=0,2=0,3=l,4=4,5=bertiniorange]{
	\pgfmathsetmacro{\t}{#4}
	\begin{scope}[shift={(#1,#2)}]
		\pgfmathsetmacro{\steps}{ceil((\t-1)/2)}
		\ifthenelse{\equal{#3}{l}}{
			\foreach \i[evaluate=\i as \x using -\t+2*\i] in {0,...,\steps}{
				\thetastate[\x][0][1][#5]
			}
		}{
			\foreach \i[evaluate=\i as \x using -\t+2*\i] in {0,...,\steps}{      
				\thetastate[-\x][0][1][#5]
			}
		}
	\end{scope}
}

\newcommandx{\drawinitstateflipped}[5][1=0,2=0,3=l,4=4,5=bertiniorange]{
	\pgfmathsetmacro{\t}{#4}
	\begin{scope}[shift={(#1,#2)}]
		\pgfmathsetmacro{\steps}{ceil((\t-1)/2)}
		\ifthenelse{\equal{#3}{l}}{
			\foreach \i[evaluate=\i as \x using -\t+2*\i] in {0,...,\steps}{
				\thetastateflipped[\x][0][1][#5]
			}
		}{
			\foreach \i[evaluate=\i as \x using -\t+2*\i] in {0,...,\steps}{      
				\thetastateflipped[-\x][0][1][#5]
			}
		}
	\end{scope}
}

\newcommandx{\eigenDL}[6][1=0,2=0,3=l,4=4,5=bertiniorange,6=topright]{
	\begin{scope}[shift={(#1,#2)}]
		\pgfmathsetmacro{\t}{#4}
		\ifthenelse{\equal{#3}{l}}{
			\eigenVL[0][0][l][\t][tr][init][#5][#6]
			\pgfmathsetmacro{\t}{#4-1}
			\rightriangle[1][0][\t][#5][#6]
			\drawinitstate[1][0][r][\t][#5]
		}{
			\begin{scope}[shift={(-0.5,0.5)}]
				\foreach \i[evaluate=\i as \x using \i, evaluate=\i as \y using \i] in {0,...,\t}{      
					\draw (\x,\y)--++(0.5,0);
					\draw[fill=white] (\x,\y) circle (0.15);
				}
			\end{scope}
		}
	\end{scope}
}

\newcommandx{\eigenDR}[6][1=0,2=0,3=l,4=4,5=bertiniorange,6=topright]{
	\begin{scope}[shift={(#1,#2)}]
		\pgfmathsetmacro{\t}{#4}
		\ifthenelse{\equal{#3}{r}}{
			\eigenVR[0][0][r][\t][tr][init][#5][#6]
			\pgfmathsetmacro{\t}{#4-1}
			\leftriangle[-1][0][\t][#5][#6]
			\drawinitstate[-1][0][l][\t][#5]
		}{
			\begin{scope}[shift={(0.5,0.5)}]
				\foreach \i[evaluate=\i as \x using \i, evaluate=\i as \y using \t-\i] in {0,...,\t}{      
					\draw (\x,\y)--++(0.5,0);
					\draw[fill=white] (\x+0.5,\y) circle (0.15);
				}
			\end{scope}
		}
	\end{scope}
}

\newcommandx{\idonpurity}[2][1=0,2=0]
{
	\begin{scope}[shift={(#1,#2)}]
		\draw[thick] (-0.5,0)--++(-0.1,0.1)--++(0,0.2)--++(0.1,-0.1);
		\draw[thick] (-0.5,0.4)--++(-0.1,0.1)--++(0,0.2)--++(0.1,-0.1);
		\draw[thick] (0.5,0)--++(0.1,0.1)--++(0,0.2)--++(-0.1,-0.1);
		\draw[thick] (0.5,0.4)--++(0.1,0.1)--++(0,0.2)--++(-0.1,-0.1);
	\end{scope}
}

\newcommandx{\swaponpurity}[2][1=0,2=0]
{
	\begin{scope}[shift={(#1,#2)}]
		\draw[thick] (-0.5,0)--++(-0.2,0.2)--++(0,0.6)--++(0.2,-0.2);
		\draw[thick] (-0.5,0.2)--++(-0.075,0.075)--++(0,0.2)--++(0.075,-0.075);
		\draw[thick] (+0.5,0)--++(+0.2,0.2)--++(0,0.6)--++(-0.2,-0.2);
		\draw[thick] (+0.5,0.2)--++(+0.075,0.075)--++(0,0.2)--++(-0.075,-0.075);
	\end{scope}
}

\newcommandx{\hook}[4][1=0,2=0,3=t,4=l]{
	\begin{scope}[shift={(#1,#2)}]
		\ifthenelse{\equal{#3}{t}}{
			\ifthenelse{\equal{#4}{l}}{\draw[thick] (0.5,-0.5) arc (45:-90:0.15);}{\draw[thick] (0.5,-0.5) arc (45:270:0.15);}
		}{\ifthenelse{\equal{#4}{l}}{\draw[ thick] (0.5,-0.5) arc (-45:90:0.15);}{\draw[ thick] (0.5,-0.5) arc (315:90:0.15);}
		}
	\end{scope}
}

\newcommandx{\hhook}[4][1=0,2=0,3=t,4=l]{
	\begin{scope}[shift={(#1,#2)}]
		\ifthenelse{\equal{#3}{t}}{
			\ifthenelse{\equal{#4}{l}}{\draw[thick] (0.5,-0.5) arc (-45:175:0.15);}{\draw[thick] (0.5,-0.5) arc (225:0:0.15);}
		}{\ifthenelse{\equal{#4}{l}}{\draw[ thick] (0.5,-0.5) arc (-45:180:-0.15);}{\draw[ thick] (0.5,-0.5) arc (45:-180:0.15);}
		}
	\end{scope}
}

\definecolor{FcolU}{rgb}{0.71,0.78,0.91}
\definecolor{colLines}{rgb}{0.31,0.31,0.31}
\definecolor{colVMPSLines}{rgb}{0.11,0.11,0.11}
\definecolor{IcolUc}{rgb}{0.71,0.41,0.42}
\definecolor{IcolU}{rgb}{0.71,0.8,0.76}
\definecolor{IcolVMPSc}{rgb}{0.73,0.69,0.7}
\definecolor{IcolVMPS}{rgb}{0.81,0.77,0.78}
\definecolor{colObs}{rgb}{1.,1.,1.}

\newcommandx{\eightlegs}[2][1=0,2=0]{
	\begin{scope}[shift={(#1,#2)}]
		\foreach \x in {1,...,8}{
			\draw (\x, 0)--++(0,0.25);
			\draw[fill] (\x,0) circle (0.05);
		}
		\foreach \x in {1,3}{
			\pgfmathsetmacro\result{2*\x-1} 
			\node () at (\result,-0.5) {$i_{\x}$};
			\pgfmathsetmacro\result{2*\x}
			\node () at (\result,-0.5) {$j_{\x}$};	
		}
		\foreach \x in {2,4}{
			\pgfmathsetmacro\result{2*\x} 
			\node () at (\result,-0.5) {$i_{\x}$};
			\pgfmathsetmacro\result{2*\x-1}
			\node () at (\result,-0.5) {$j_{\x}$};	
		}
	\end{scope}
}

\usepackage{hyperref}

\theoremstyle{plain}

\newcommand{\be}{\begin{equation}}
\newcommand{\ee}{\end{equation}}
\definecolor{mygreen}{rgb}{0.0,0.55,0.3}

\newcommand{\M}{\mathcal{M}}
\newcommand{\tU}{\tilde{U}}

\newcommand{\er}[1]{Eq.~\eqref{#1}}

\newcommand{\Er}[1]{Equation~\eqref{#1}}

\begin{document}

\title{Efficient post-selection in light-cone correlations of monitored quantum circuits}
\author{Jimin Li}
\affiliation{Department of Applied Mathematics and Theoretical Physics, University of Cambridge, Wilberforce Road, Cambridge CB3 0WA, United Kingdom}
\author{Robert L. Jack}
\affiliation{Department of Applied Mathematics and Theoretical Physics, University of Cambridge, Wilberforce Road, Cambridge CB3 0WA, United Kingdom}
\affiliation{Department of Chemistry, University of Cambridge,
Lensfield Road, Cambridge CB2 1EW, United Kingdom}
\author{Bruno Bertini}
\affiliation{School of Physics and Astronomy, University of Birmingham, Birmingham B15 2TT, United Kingdom}
\author{Juan P. Garrahan }
\affiliation{School of Physics and Astronomy, University of Nottingham, Nottingham, NG7 2RD, UK}
\affiliation{Centre for the Mathematics and Theoretical Physics of Quantum
Non-Equilibrium Systems, University of Nottingham, Nottingham, NG7 2RD, UK}

\date{\today}

\begin{abstract}
We consider how to target evolution conditioned on atypical measurement outcomes in monitored quantum circuits, i.e., the post-selection problem. We show that for a simple class of measurement schemes, post-selected light-cone dynamical correlation functions can be obtained efficiently from the averaged correlations of a different unitary circuit. This connects rare measurement outcomes in one circuit to typical outcomes in another one. We derive conditions for the existence of this rare-to-typical mapping in brickwork quantum circuits made of XYZ gates. We illustrate these general results with a model system that exhibits a dynamical crossover (a smoothed dynamical transition) in event statistics, and discuss extensions to more general dynamical correlations. 
\end{abstract}

\maketitle

\section{Introduction}

For closed quantum many-body systems consisting of a subsystem $A$ and its complement $B$, there are various ways to describe the dynamical evolution~\cite{dalessio2016from,breuer2002the-theory,gardiner2004quantum}. The most fundamental is the coherent evolution of the total state $\ket{\Psi_{A \cup B}}$ through a unitary operator acting on $A \cup B$. Discarding all information about $B$, one can follow the evolution of the reduced state $\rho_A = \Tr_B \ket{\Psi_{A \cup B}}\bra{\Psi_{A \cup B}}$, with its dissipative dynamics given by a quantum channel $\Phi$, which is in general non-Markovian~\cite{breuer2002the-theory,gardiner2004quantum}. When $\Phi$ is given (either exactly or approximately) by a Markovian channel $\M$~\cite{breuer2002the-theory,gardiner2004quantum,lindblad1976on-the-generators,gorini1976completely}, there is a third, intermediate, level of detail where some information about $B$ is retained. This is the ``unravelling'' of $\M$ into {\em quantum trajectories}~\cite{breuer2002the-theory,gardiner2004quantum}: $B$ is measured, and the evolution of the state on $A$ is conditioned on the measurement outcomes, producing stochastic dynamics~\cite{belavkin1990a-stochastic,dalibard1992wave-function,plenio1998the-quantum-jump}. Averaging over the trajectories recovers the channel $\M$.  

The estimation of expectation values along specific trajectories or sets of trajectories is known as {\em post-selection}~\cite{fisher2023random,garratt2024probing,mcginley2024postselection-free,passarelli2024many-body} and amounts to targeting {\it rare events} in the dynamics. This gives much more information compared to the mere average, including time-correlations to all orders, and the probabilities of atypical occurrences~\cite{garrahan2018aspects}. Unfortunately, it is very challenging to achieve post-selection in practical experiments~\cite{koh2023measurement-induced,choi2023preparing} because of the inherent stochasticity of quantum measurement. The estimation of any expectation value already requires many samples, but post-selection is much worse because each individual sample is exponentially rare, and one still requires large numbers of them. This introduces a cost that scales exponentially in time, and sometimes also in space. 

Here we present a general method to address the post-selection problem in monitored quantum circuits by directly accessing evolution conditioned on atypical measurement outcomes. Quantum circuits have become a central platform for studying quantum dynamics (see e.g.~\cite{nahum2017quantum,nahum2018dynamics,nahum2018operator,chan2018solution,keyserlingk2018operator,bertini2019entanglement,bertini2019exact,gopalakrishnan2019unitary,friedman2019spectral,rakovszky2019sub-ballistic,piroli2020exact, claeys2020maximum, bertini2020scrambling, klobas2021exact, bertini2023localised,fisher2023random, lorenzo-triviality-23, wang24-arxiv,cech24-arxiv}), and in this setting the post-selection problem is directly relevant to the highly debated measurement-induced phase transitions~\cite{li2019measurement-driven,skinner2019measurement-induced,zabalo2020critical}. Through exact analytical calculations we show that for a class of measurement schemes in ``brickwork'' circuits the post-selected dynamics along the light-cone can be obtained efficiently from the dynamics of another brickwork circuit. This connects rare measurement outcomes in a circuit to typical outcomes in a different (auxiliary) circuit {\em in the same class}. That is, both circuits are defined on the same Hilbert space, with the same interaction range, but no post-selection is required in the auxiliary circuit. 
We derive conditions for the existence of this {\em rare-to-typical mapping} and illustrate our results with an example that exhibits a dynamical crossover (a smoothed transition due to finite size) in the measurement statistics. We finish by discussing extensions to more general scenarios.

\begin{figure}[t]
\centering
\begin{tikzpicture}[baseline={([yshift=-0.6ex]current bounding box.center)},scale=.5]
\draw[<->,black, thick] (0.5,-0.25)--(5.5,-.25);
\draw[<->,black, thick] (6,1.25)--(6,6.5);
\node at (3,-1) {$2L$};
\node at (6.5,3) {$2t$};
\node at (-.5,6.5) {(a)};
\foreach \i in {-2,...,0}
{
\draw[gray, dotted] (-.5,2*\i+5.25)--(5.5,2*\i+5.25);
\draw[gray, dotted] (-.5,2*\i+4.75)--(5.5,2*\i+4.75);
}
\foreach \i in {0,...,2}{
\draw[fill=black] (2*\i+.5,0.5) circle (0.1);
\draw[fill=black] (2*\i+1.5,0.5) circle (0.1);
\tsfmatV[2*\i][0][r][6][][][bertinired][topright]}
\foreach \i in {2,4,6}{
\hook[5][\i][t][l]
\hook[5][\i-1][b][l]
\hook[-1][\i][t][r]
\hook[-1][\i-1][b][r]}
\end{tikzpicture}
\quad\qquad
\begin{tikzpicture}[baseline={([yshift=-0.6ex]current bounding box.center)},scale=.5]
\node at (-.5,6.5) {(b)};
\node at (1.5,-.25) {$x=0$};
\foreach \i in {-2,...,0}
{
\draw[gray, dotted] (-.5,2*\i+5.25)--(5.5,2*\i+5.25);
\draw[gray, dotted] (-.5,2*\i+4.75)--(5.5,2*\i+4.75);
}
\foreach \i in {0,...,2}{
\draw[fill=black] (2*\i+.5,0.5) circle (0.1);
\draw[fill=black] (2*\i+1.5,0.5) circle (0.1);
\tsfmatV[2*\i][0][r][6][][][bertinired][topright]}
\foreach \i in {2,4,6}{
\hook[5][\i][t][l]
\hook[5][\i-1][b][l]
\hook[-1][\i][t][r]
\hook[-1][\i-1][b][r]}
\foreach \i in {0,...,5}{
\draw (.5+\i,.5+\i) pic[rotate = 45, red] {cross=3pt};
}
\foreach \i in {0,...,3}{
\draw (.5+\i+2,.5+\i) pic[rotate = 45, red] {cross=3pt};
}
\draw (.5-2+2,.5+4) pic[rotate = 45, red] {cross=3pt};
\draw (.5-1+2,.5+5) pic[rotate = 45, red] {cross=3pt};
\draw (.5-1+3,.5+6) pic[rotate = 45, red] {cross=3pt};
\draw (.5-1+1,.5+6) pic[rotate = 45, red] {cross=3pt};
\end{tikzpicture}
\begin{tikzpicture}[baseline={([yshift=-0.6ex]current bounding box.center)},scale=.5]
\node at (5.7,4.25) {$a$};
\node at (1.15,.5) {$b$};
\node at (-.5,4.5) {(c)};
\foreach \i in {-2,...,-1}
{
\draw[gray, dotted] (-.5,2*\i+5.25)--(5.5,2*\i+5.25);
\draw[gray, dotted] (-.5,2*\i+4.75)--(5.5,2*\i+4.75);
}
\foreach \i in {-3.575,...,-2.575}
{
\draw[gray, dotted] (-.5,2*\i+5.25)--(5.5,2*\i+5.25);
\draw[gray, dotted] (-.5,2*\i+4.75)--(5.5,2*\i+4.75);
}
\foreach \i in {0,...,2}{
\tsfmatV[2*\i][0][r][4][][][bertinired][topright]}
\foreach \i in {2,4}{
\hook[5][\i][t][l]
\hook[5][\i-1][b][l]
\hook[-1][\i][t][r]
\hook[-1][\i-1][b][r]}
\foreach \i in {0,...,2}{
\tsfmatV[2*\i][-4.15][l][4][][][bertiniblue][topright]}
\foreach \i in {1-.15,1-2.15}{
\hook[5][\i][t][l]
\hook[5][\i-1][b][l]
\hook[-1][\i][t][r]
\hook[-1][\i-1][b][r]}
\draw[fill=black] (2*1-.5,0.425) circle (0.1);
\draw[fill=black] (2*2+1+.5,4.5) circle (0.1);
\foreach \i in {0,2,4}{
\hhook[\i][-3.15][b][l]
\hhook[\i+1][-3.15][b][r]
\hhook[\i+1][5][t][l]
\hhook[\i][5][t][r]}
\foreach \i in {0,...,2}
{
\draw[black, thick] (2*\i+.5,0.34)--(2*\i+.5,0.52);
\draw[black, thick] (2*\i+1.5,0.34)--(2*\i+1.5,0.52);
}
\foreach \i in {0,...,4}{
\draw[fill=red] (.5+\i,.5+\i) circle (0.08);
}
\foreach \i in {0,...,3}{
\draw[fill=red] (.5+\i+2,.5+\i) circle (0.08);
}
\foreach \i in {0,...,4}{
\draw[fill=red] (.5+\i,.5-\i-0.15) circle (0.08);
}
\foreach \i in {0,...,3}{
\draw[fill=red] (.5+\i+2,.5-\i-0.15) circle (0.08);
}
\draw[fill=red] (.5,.5-4-0.15) circle (0.08);
\draw[fill=red] (.5,.5+4) circle (0.08);
\node[below, scale=0.7] at (.5+0+2,.5-0-0.15)  {$0$};
\node[below, scale=0.7] at (.5+1+2,.5-1-0.15)  {$1$};
\node[below, scale=0.7] at (.5+2+2,.5-2-0.15)  {$1$};
\node[below, scale=0.7] at (.5+3+2,.5-3-0.15)  {$0$};
\node[below, scale=0.7] at (.5+0,.5-0-0.15)  {$1$};
\node[below, scale=0.7] at (.5+1,.5-1-0.15)  {$1$};
\node[below, scale=0.7] at (.5+2,.5-2-0.15)  {$0$};
\node[below, scale=0.7] at (.5+3,.5-3-0.15)  {$0$};
\node[above, scale=0.7] at (.5+4,.5-4-0.15)  {$0$};
\node[above, scale=0.7] at (.5,.5-4-0.15)  {$1$};

\node[above, scale=0.7] at  (.5+0,.5+0) {$1$};
\node[above, scale=0.7] at  (.5+1,.5+1) {$1$};
\node[above, scale=0.7] at (.5+2,.5+2) {$0$};
\node[above, scale=0.7] at  (.5+3,.5+3) {$0$};
\node[below, scale=0.7] at  (.5+4,.5+4) {$0$};
\node[below, scale=0.7] at  (.5,.5+4) {$1$};

\node[above, scale=0.7] at  (.5+0+2,.5+0) {$0$};
\node[above, scale=0.7] at  (.5+1+2,.5+1) {$1$};
\node[above, scale=0.7] at (.5+2+2,.5+2) {$1$};
\node[above, scale=0.7] at  (.5+3+2,.5+3) {$0$};
\end{tikzpicture}
\quad\quad\qquad	
\begin{tikzpicture}[baseline={([yshift=-0.6ex]current bounding box.center)},scale=.5]
\node at (5.75,4.25) {$a$};
\node at (1.15,.5) {$b$};
\node at (1.5,4.5) {(d)};
\foreach \i in {1,...,4}{
\draw[gray, dotted] (\i+.75,\i+.5)--(\i+.75,-\i+0.15);
}
\draw[gray, dotted] (4.5+.75,4+.5)--(4.5+.75,-4+0.15);
\foreach \i in {0,...,3}{
\tsfmatV[\i+2][\i][r][1][][][bertinired][topright]}
\foreach \i in {0,...,3}{
\tsfmatV[\i+1][-\i-1.15][l][1][][][bertiniblue][topright]}
\draw[fill=black] (2*1-.5,0.425) circle (0.1);
\draw[fill=black] (2*2+1+.5,4.5) circle (0.1);
\foreach \i in {1,...,4}{
\hhook[\i][-.15-\i+1][b][l]
\hhook[\i][\i+1][t][r]
}
\foreach \i in {1,...,2}
{
\draw[black, thick] (2*\i+.5,2*\i-1.5)--(2*\i+.5,-2*\i-.15+2.5);
\draw[black, thick] (2*\i+1.5,2*\i-.5)--(2*\i+1.5,-2*\i-.15+1.5);
}
\hhook[4+1][-.15-4+1][b][r]
\hhook[4+1][5][t][l]
\foreach \i in {1,...,4}{
\draw[fill=red] (.5+\i,.5+\i) circle (0.08);
}
\foreach \i in {0,...,3}{
\draw[fill=red] (.5+\i+2,.5+\i) circle (0.08);
}
\foreach \i in {1,...,4}{
\draw[fill=red] (.5+\i,.5-\i-0.15) circle (0.08);
}
\foreach \i in {0,...,3}{
\draw[fill=red] (.5+\i+2,.5-\i-0.15) circle (0.08);
}
\node[below, scale=0.7] at (.5+0+2,.5-0-0.15)  {$0$};
\node[below, scale=0.7] at (.5+1+2,.5-1-0.15)  {$1$};
\node[below, scale=0.7] at (.5+2+2,.5-2-0.15)  {$1$};
\node[below, scale=0.7] at (.5+3+2,.5-3-0.15)  {$0$};
\node[above, scale=0.7] at (.5+1,.5-1-0.15)  {$1$};
\node[above, scale=0.7] at (.5+2,.5-2-0.15)  {$0$};
\node[above, scale=0.7] at (.5+3,.5-3-0.15)  {$0$};
\node[above, scale=0.7] at (.5+4,.5-4-0.15)  {$0$};

\node[below, scale=0.7] at  (.5+1,.5+1) {$1$};
\node[below, scale=0.7] at (.5+2,.5+2) {$0$};
\node[below, scale=0.7] at  (.5+3,.5+3) {$0$};
\node[below, scale=0.7] at  (.5+4,.5+4) {$0$};

\node[above, scale=0.7] at  (.5+0+2,.5+0) {$0$};
\node[above, scale=0.7] at  (.5+1+2,.5+1) {$1$};
\node[above, scale=0.7] at (.5+2+2,.5+2) {$1$};
\node[above, scale=0.7] at  (.5+3+2,.5+3) {$0$};
\end{tikzpicture}
\caption{Diagrammatic representation of the quantum circuit. Red boxes represent gate $U$ and blue ones $U^\dag$; each wire is associated to the Hilbert space of a qubit and connected wires indicate a sum over their internal states. (a)~State at time $t=3$ for $2L=6$ qubits. (b)~Measurement scheme, with red crosses denoting measurements. (c)~Light-cone correlation function for a given quantum trajectory, red dots denote projectors $P^{(i)}$ with $i=0$ or $i=1$ at each half time step, depending on the trajectory. (d)~Simplified light-cone correlation function, using unitarity. }
\label{fig:TNdiagrams}
\end{figure}
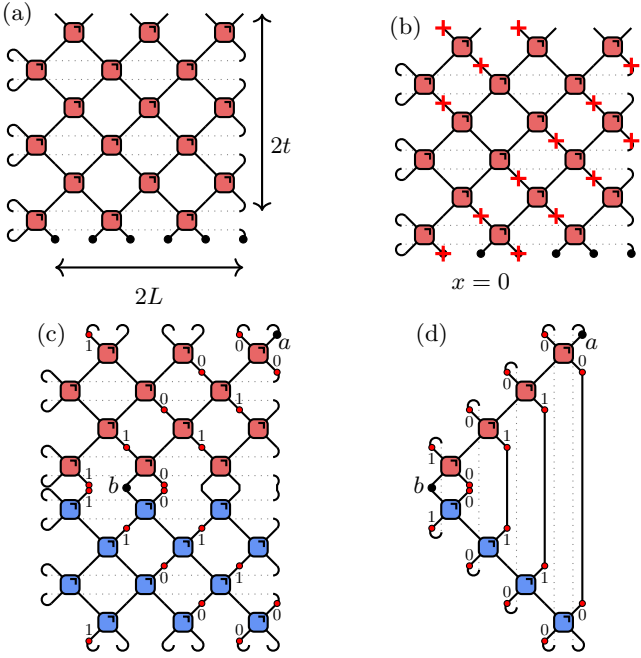

\begin{figure}[t]
  \centering
  \includegraphics[width=0.9\columnwidth]{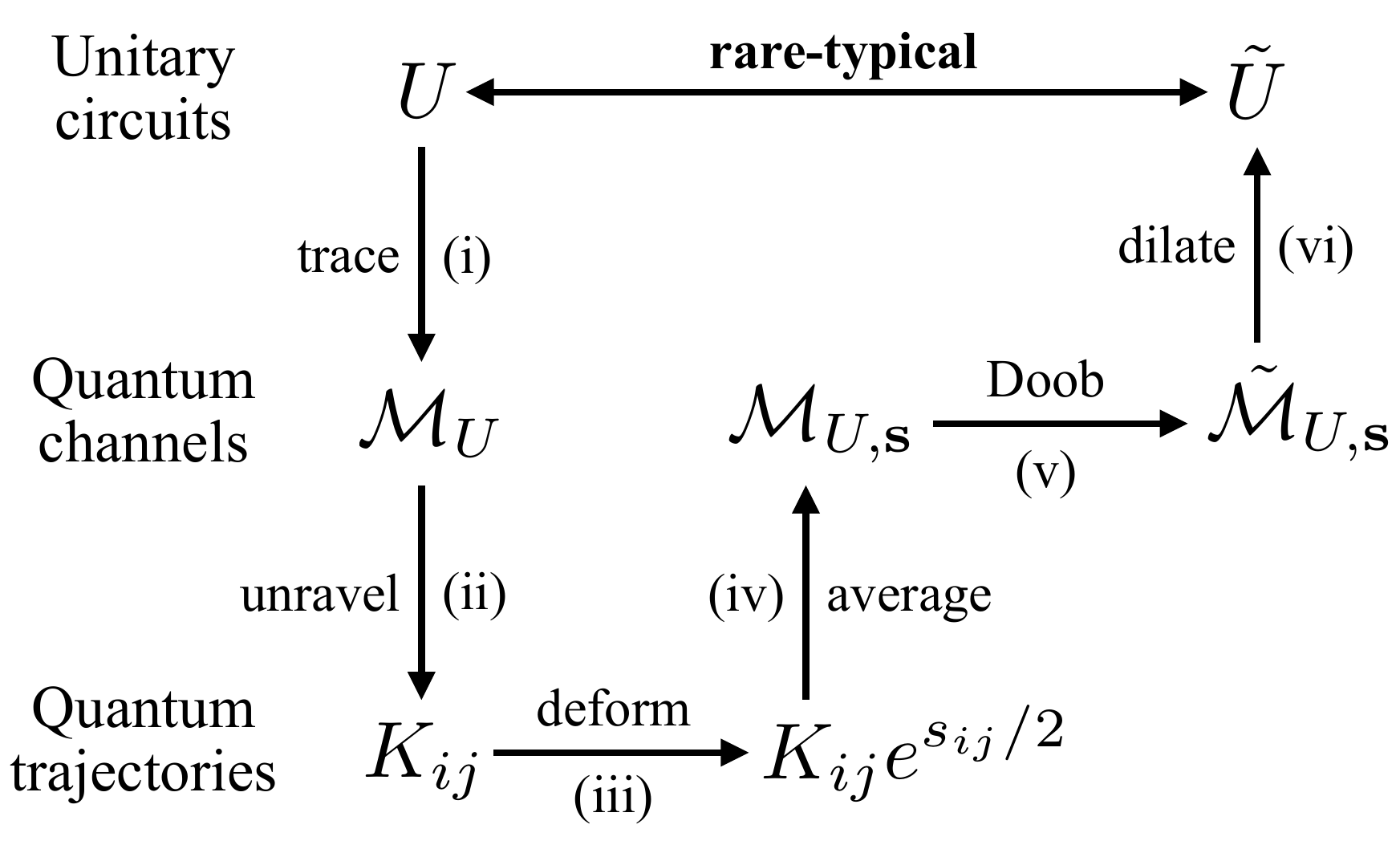}
  \caption{Post-selecting rare trajectories of one circuit as typical trajectories of another.
Tracing over a brickwork circuit (i) generates a Markovian channel $\M$. which is unravelled (ii) to yield Kraus operators $\{ K\}$ and quantum trajectories.  Post-selection is achieved by reweighting the trajectories (iii) or equivalently by tilting the channel with a biasing field $s$ (iv).  The quantum Doob transformation (v) yields a new channel $\tilde{\M}_{\boldsymbol s}$ which optimally generates the post-selected (rare) trajectories.  Finally, a dilation of this channel (vi) provides a new brickwork circuit with gate $\tilde{U}$.
  }  
  \label{fig:cycle}
\end{figure}

\section{setup}

Our illustrative example is based on a well-studied one-dimensional unitary brickwork quantum circuit, i.e., a chain of $2L$ qubits evolved in time by the staggered application of two-qubit unitary gates~\cite{fisher2023random}. The system state at time $t\in \mathbb N$ is $\ket{\Psi_t}= \mathbb{U}^t \ket{\Psi_0}$ where $\ket{\Psi_0}$ is the initial state and 
\be
\mathbb{U} =\bigotimes_{x=1}^{L} U_{x-1/2,x} \bigotimes_{x=1}^{L} U_{x,x+1/2}\,.
\label{eq:U}
\ee
We label sites by half integers, use periodic boundary conditions, and denote by $U_{x,y}$ the operator acting as the two-qubit gate $U\in {\rm U}(4)$ on the qubits at positions $x$ and $y$ (and the identity elsewhere). Fig.~\ref{fig:TNdiagrams}(a) shows the circuit in the standard diagrammatic representation of tensor networks~\cite{cirac2021matrix}, which greatly facilitates the analytical analysis. A self-contained review of the latter is presented in Appendix~\ref{sec:diagram}.

We now implement the following measurement scheme: every half time step $\tau\in[0,t]\cap\mathbb Z/2$ 
we measure in the computational basis $\{\ket{0}, \ket{1}\}$ the qubits at spatial positions $\tau \pm 1/2$ (i.e., along the direction of the light-cone), see Fig.~\ref{fig:TNdiagrams}(b). These measurements make the evolution stochastic. Denoting by $\ket{\Psi_\tau}$ the state of the system at $\tau$, the probability of measuring at time $\tau+1/2$ the qubit at position $\tau$ in the state $\ket{i}$ and that at position $\tau+1$ in the state $\ket{j}$ is given by  $\|\mathbb K_{ij}(\tau)\ket{\Psi_{\tau}}\|^2$, where the many-body Kraus operators are
\be
\!\!\!\!\mathbb K_{ij}(\tau) \!=\!
\begin{cases}
\displaystyle P^{(i)}_{\tau} P^{(j)}_{\tau+1} \bigotimes_{x=1}^{L} U_{x-1/2,x} & \tau \in \mathbb Z+{1}/{2}\\
\\
\displaystyle P^{(i)}_{\tau} P^{(j)}_{\tau+1} \bigotimes_{x=1}^{L} U_{x,x+1/2} & \tau \in \mathbb Z
\end{cases}\!\!,
\label{eq:KrausMB}
\ee
and $P^{(i)}$ denotes a projector on the state $\ket{i}$ of the computational basis. The (normalised) state of the system after the measurement is then $\mathbb K_{ij}(\tau)\ket{\Psi_{\tau}}/\|\mathbb K_{ij}\ket{\Psi_{\tau}}\|$. A full measurement sequence generates the quantum trajectory $(\boldsymbol i, \boldsymbol j) = (i_0, \ldots, i_{2t}, j_0, \ldots, j_{2t})$ occurring with probability  $\|\mathbb K_{\boldsymbol i \boldsymbol j}(t)\ket{\Psi_{0}}\|^2$, where $\mathbb K_{\boldsymbol i \boldsymbol j}(t)\equiv \mathbb K_{i_{2t}j_{2t}}(t-1/2)\cdots \mathbb K_{i_0 j_0}(-1/2)$~\footnote{We set $K_{i j}(-1/2)=P^{(i)}_{-1/2} P^{(j)}_{1/2}$.}. 

Consider now the infinite temperature correlation function for one qubit along the light-cone passing through $x=0$. For a given trajectory this reads
\be
\!\!C^{ab}_{(\boldsymbol i, \boldsymbol j)}(t)\!= \frac{1}{2^{2L}} \sum_{\boldsymbol s} \expval{\mathbb K_{\boldsymbol i \boldsymbol j}^{\dag}(t)  a_t \mathbb K^{\phantom{\dag}}_{\boldsymbol i \boldsymbol j}(t) b_0}{\boldsymbol s},
\label{eq:correlationgen}
\ee
where $\{\ket{\boldsymbol s}\}$ is a basis of the Hilbert space and we used the notation $a_x$ to denote the operator that acts non-trivially, as $a\in {\rm End}(\mathbb C^2)$, only at site $x$. The correlation function \eqref{eq:correlationgen} is a prototypical example of an observable ($a_t$) measured on a post-selected trajectory ($\boldsymbol{i},\boldsymbol{j}$). 
We find that it displays complex rare-event behavior, while remaining simple enough for exact analysis.

In graphical form Eq.~\eqref{eq:correlationgen} is given in Fig.~\ref{fig:TNdiagrams}(c). This tensor network can be drastically simplified because the causal light-cones of the two operators overlap only on one line in the discrete space time (cf.\ Refs.~\cite{bertini2019exact, gutkin2020local, gutkin2020exact, kos2021correlations}). For $L>2t+1$ this yields Fig.~\ref{fig:TNdiagrams}(d), or in formulae, 
\be
\!\!C^{ab}_{(\boldsymbol i, \boldsymbol j)}(t)\!=\!\tr[a K^{\phantom{\dag}}_{i_{2t} j_{2t-1} } \!\!\cdots K^{\phantom{\dag}}_{i_1j_0 } b K^\dag_{i_1j_0 } \!\!\cdots K^\dag_{i_{2t} j_{2t-1} } ],
\label{eq:correlationtrajectory}
\ee
where the {\em single site Kraus operators} are
\be
\mel{m}{K_{ij}}{n}= \frac{1}{\sqrt{2}}  \mel{i m}{U}{n j} \, .
\label{eq:Kloc}
\ee
Summing over all trajectories, we recover the dynamical correlator along the light-cone of the unmeasured circuit~\cite{bertini2019exact, gutkin2020local, gutkin2020exact, kos2021correlations}
\be
\sum_{\boldsymbol i, \boldsymbol j} C^{ab}_{(\boldsymbol i, \boldsymbol j)}(t) = C^{ab}(t) = \tr[a \mathcal{M}^{2t}_{U}[b]],
\label{eq:sumoveroutcomes}
\ee 
where 
 \be
\mathcal{M}_{U}[\rho] = \frac{1}{2} \tr_A[ U (\rho \otimes \mathbf{1}_B) U^\dag],
\label{eq:channel}
 \ee 
is the quantum channel for the average dynamics of a single qubit along the light-cone,
with $\tr_A[\cdot]$ the trace over the first site. \Er{eq:sumoveroutcomes} holds because the channel is ``unravelled''~\cite{bengtsson2007geometry} by the $\{K_{ij}\}$ 
\footnote{
In general Kraus operators only need to obey 
$\sum_{ij} K_{ij}^\dag K_{ij} = \mathbf{1}$. Note however that in our case we have 
$\sum_{i} K_{ij}^\dag K_{ij} = \mathbf{1}/2$ for all $j$, as a consequence of the local unitary $U$ in \er{eq:Kloc} and of the brickwork structure of the circuit. 
}
\be
\mathcal{M}_{U}[\cdot] = \sum_{i,j} K_{ij} \cdot K^\dag_{ij} . 
\ee

\section{Reweighting of trajectories and large deviations.}
We aim to characterise correlations along quantum trajectories in a way that sidesteps the post-selection problem. We follow the protocol of Fig.~\ref{fig:cycle}, in which steps (i) and (ii) correspond to the derivation of $\mathcal{M}_{U}$ and $K_{ij}$ above.
We now execute steps (iii) to (v) (see Appendices \ref{sec:light-cone-channel} and \ref{sec:unravelling} for details).

Instead of focusing on a specific trajectory, we consider a ``soft" post-selection of all trajectories characterised by a certain pattern of measurement outcomes in the limit of large time $t$.  This is amenable to quantum large deviation methods (LD) \cite{esposito2009nonequilibrium,garrahan2010thermodynamics}.  Starting from the sum over measurement outcomes in Eq.~\eqref{eq:sumoveroutcomes}, we (exponentially) reweigh some terms with respect to others. The resulting biased ensemble of trajectories is encoded in a deformed (or ``tilted'') quantum channel ${\mathcal M}_{U,s}$~\cite{garrahan2010thermodynamics}. 

Consider the trajectory determined by the sequence $((i_0, j_1),\ldots, (i_{2t-1}, j_{2t}))$, where $(i_k,j_{k+1})$ denotes a binary pair of outcomes for $k=1,\ldots,2t-1$ (cf.\ the configuration of a classical Ising model on a ladder, where $k$ denotes a rung). We reweigh the trajectories based on how many rungs have outcome pairs $(0,0)$, $(1,0)$, $(0,1)$, and $(1,1)$. This is achieved by introducing four {\em counting fields} 
$\boldsymbol s= \{s_{ij}\}_{i,j=0,1}$ and modifying the sum in Eq.~\eqref{eq:sumoveroutcomes},
\be
\sum_{\boldsymbol i, \boldsymbol j} e^{- \sum_{l,m} s_{lm} Q_{lm}(\boldsymbol i,\boldsymbol j)}C^{ab}_{(\boldsymbol i, \boldsymbol j)}(t,t) \equiv \tr[a \mathcal{M}^{2t}_{U,\bold{s}}[b]]
\label{eq:biasedsum}
\ee
where $Q_{lm}(\boldsymbol i,\boldsymbol j)= \sum_{k=1}^{2t} \delta_{i_{k-1},l}\delta_{j_{k},m}$. Since the sums on each rung are independent the tilted channel reads \cite{garrahan2010thermodynamics}
\begin{equation}
    \mathcal{M}_{U,\bold{s}}[\rho] = \sum_{i,j} e^{-{s}_{ij}} K_{ij} \rho K^{\dagger}_{ij}.
    \label{eq:deformedchannel}
\end{equation}
This result implements steps (iii) and (iv) of Fig.~\ref{fig:cycle}.

Note that since the sum of all $Q_{\mu\nu}(\boldsymbol i,\boldsymbol j)$ is fixed to $2t-1$, the bias $\boldsymbol{s}$ corresponds to three non-trivial fields.
The exponential tilting in Eq.~\eqref{eq:deformedchannel} controls the average number of outcomes of each type, instead of their exact values, but these ``soft-constrained'' (canonical) trajectory ensembles are equivalent to ``hard-constrained'' (microcanonical) ensembles, for large times~\cite{chetrite2013nonequilibrium}.

The operator $\mathcal{M}_{U,\bold{s}}$ is not trace-preserving [its leading eigenvalue $e^{\theta(\bold{s})}\neq 1$ in general], so it is not a physical quantum map.
Step (v) of Fig.~\ref{fig:cycle} performs the quantum version~\cite{garrahan2010thermodynamics, carollo2018making, carollo2019unraveling, cech2023thermodynamics} of a {\em generalised Doob transform} \cite{jack2010large,chetrite2015nonequilibrium}, yielding a new trace-preserving channel which reproduces the biased trajectory ensemble. Specifically 
\be
    \!\!\!\!\!\!\!\!\tilde{\mathcal{M}}_{U,\bold{s}}[\cdot] \!\!=\!\! \sum_{i,j} \tilde{K}_{ij}\! \cdot\! \tilde{K}_{ij}^{\dagger},
    \quad
    \tilde{K}_{ij} \!=\! \frac{e^{-s_{ij}/2}}{e^{\theta(\bold{s})/2} } l^{1/2}_{\bold{s}} K_{ij} l^{-1/2}_{\bold{s}}\!\!\!,
    \label{eq:Doob}
\ee
with $l_{\bold{s}}$ the leading left eigenmatrix of $\mathcal{M}_{U,\bold{s}}$ (see Appendix \ref{sec:largde_deviation} for details). In terms of $\tilde{\mathcal{M}}_{U,\bold{s}}$, the reweighted trajectories \eqref{eq:biasedsum} are 
\be
\!\!\!\!\tr[a \mathcal{M}^{2t}_{U,\bold{s}}[b]]
 \!\!=\! e^{2 t\theta(\bold{s})} \!\tr[l^{-1/2}_{\bold{s}} a l^{-1/2}_{\bold{s}} \tilde{\mathcal{M}}^{2t}_{U,\bold{s}}[l^{1/2}_{\bold{s}} b l^{1/2}_{\bold{s}}]]\!.
\label{eq:biasedsumDoob}
\ee
Hence, typical trajectories of the channel \eqref{eq:Doob} reproduce the rare trajectories of the original $\M_U$, as encoded in the tilted $\mathcal{M}_{U,\bold{s}}$.

\subsection{Unitary circuit realising the rare events} 

We now turn to step (vi) in Fig.~\ref{fig:cycle}, which is motivated by the question:
\begin{itemize}
  \item Can one find an $\bold{s}$-dependent unitary operator $\tU\in{\rm End}(\mathbb C^2\otimes \mathbb C^2)$ such that $\tilde{\mathcal{M}}_{U,\bold{s}}$ is given by \er{eq:channel} with $U$ replaced by $\tU$? 
\end{itemize}
We emphasise that whenever the answer to this question is affirmative the post selection problem is \emph{solved}: rare quantum trajectories can be targeted with the same cost as typical ones; equivalently, hybrid quantum dynamics can be studied with the same complexity as unitary dynamics. Below we present cases where the answer is affirmative for the specific protocol in Eq.~\eqref{eq:KrausMB} and the observable in Eq.~\eqref{eq:correlationgen}. Note, however, that the same question can be asked for more general settings.



One might expect that $\tU$ can always be found, because \er{eq:channel} is an environmental representation $\M_{\tU}$~\cite{bengtsson2007geometry}.  The Stinespring dilation theorem~\cite{stinespring1955positive} guarantees existence of some environmental representation, but our question is more restrictive because (a) both system and environment in Eq.~\eqref{eq:channel} must be isomorphic to $\mathbb C^2$ (i.e., the dilation defines a unitary circuit similar to the original one); (b) the environment has to be in the maximally mixed state (i.e., we seek a \emph{unistochastic} channel~\cite{bengtsson2007geometry}).  In fact, \er{eq:channel} already implies that $\M_{\tU}$ is unital [$\M_{\tU}(\mathbf{1})=\mathbf{1}$], so existence of $\tU$ requires that $\tilde{\mathcal{M}}_{U,\bold{s}}$ is also unital, which is not the case in general.  Hence, while steps (i-v) in Fig.~\ref{fig:cycle} can be performed for any $U$ and $\bold{s}$, step (vi) is restricted to specific cases of the post-selection problem.

\subsection{Example: XYZ gates}

To illustrate this, we focus on circuits where $U$ is an XYZ gate: $U=U_{\rm{XYZ}}(\{J_i\})$, with  
\be
U_{\rm{XYZ}}(\{J_i\}) = e^{-i ({J_x} \sigma_x \otimes \sigma_x + {J_y} \sigma_{y}\otimes \sigma_{y} + {J_z} \sigma_z \otimes\sigma_z)/2} \!,
\label{eq:Heisenberg}
\ee
where $\{\sigma_{i}\}_{i=x,y,z}$ are Pauli matrices and $(J_x,J_y,J_z) \in [-\pi,\pi]$. 
This implies a $\mathbb{Z}_2$ symmetry $\tilde{\mathcal{M}}_{U,\bold{s}}[\sigma_z(\cdot)\sigma_z]=\sigma_z\tilde{\mathcal{M}}_{U,\bold{s}}(\cdot)\sigma_z$ for all $\bold{s}$, which relies on the fact that $\sigma_z$ is diagonal in the measurement basis of Eq.~\eqref{eq:KrausMB}. To replicate this, we also seek $\tilde{U}$ in XYZ form with $\bold{s}$-dependent couplings $\{ J'_{\bold{s},i}\}_{i=x,y,z}$, i.e., we search for new circuits with exactly the same complexity as the initial ones.  This choice of $\tU$ implies the additional symmetry ${\mathcal{M}}_{\tU}[\sigma_x(\cdot)\sigma_x]=\sigma_x{\mathcal{M}}_{\tU}(\cdot)\sigma_x$. 

Writing explicitly the dependence of the channels on the couplings, finding $\tU$ amounts to solving
\be
\tilde{\mathcal{M}}_{U,\boldsymbol{s}}(\{J_i\}) ={\mathcal{M}}_{\tU}(\{J'_{\bold{s},i}\}) \,,
\label{eq:channelconstraint}
\ee
for real couplings $\{J'_{\bold{s},i}\}$. Note that the symmetries of ${\mathcal{M}}_{\tU}$ restrict solutions
to cases where $\tilde{\mathcal{M}}_{\boldsymbol{s}}$ is both unital and $\mathbb{Z}_2$-symmetric. 

As shown explicitly in Appendix~\ref{sec:Doob_channel}, Eq.~\eqref{eq:channelconstraint} can be solved analytically revealing several classes of solutions. 
The first class is when any two of the $J_i$ are equal to $\pm\pi/2$, making $U$ dual unitary~\cite{bertini2019exact,gopalakrishnan2019unitary}.  For example   
\be
U_{\rm{XYZ}}(\{\tfrac \pi 2, \tfrac \pi 2, J_j \})  = S e^{ - i (J_j-\pi/2) (\sigma_{j}\otimes \sigma_{j})/2},
\label{eq:du}
\ee
where $J_j$ is the coupling not equal to $\pi/2$ and $S$ is the SWAP gate.
In these cases a dual-unitary $\tU$ also exists of the form \eqref{eq:du}. 
These post-selection problems are simple because $K_{ij} K^\dag_{ij} = c_{ij} \mathbf{1}$ for all $i,j$ (with $c_{ij}\geq0$). 
This means that the channel $\M$ is a {\em classical mixture} of unitary channels, a measurement $(i_{k-1}, j_k)$ in $C_{(\boldsymbol{i},\boldsymbol{j})}$ occurs with probability $c_{i_{k-1}, j_k}$ independent of other measurements, and the count statistics is multinomial. This simplicity is evident in the fact that $l_{\boldsymbol{s}}=\mathbf{1}$, 
making the Doob transform \eqref{eq:Doob}
straightforward, with $\tilde{\mathcal{M}}_{U,\boldsymbol{s}}$ a different mixture of the same unitaries as in $\M$. This result establishes an interesting connection to random unitary circuits \cite{fisher2023random}: atypical measurement outcomes in (space-time) translation invariant dual unitary circuits are equivalent to the evolution under atypical sequences of i.i.d.\ random unitaries along the light-cone. 

The other classes of solution to \er{eq:channelconstraint}
restrict the counting fields to $s_{00}=s_{11}$, ensuring that $\tilde{M}_{\boldsymbol{s}}$ is unital (see Appendix~\ref{sec:Doob_channel}). This means that $(0,0)$ measurement pairs cannot be selected preferentially over $(1,1)$.  The $\mathbb{Z}_2$ symmetry requires either ({\bf A}) $s_{01} = s_{10}$; or ({\bf B}) $\sin J_x = \sin J_y$; or ({\bf C}) $J_z=\pm\pi/2$.
Post-selection is simple in case ({\bf A}) because the counting fields only differentiate the outcomes with $i_{k-1}=j_k$ from those with $i_{k-1}\neq j_k$;  these two possibilities are again independent for all $k$, with $l_{\boldsymbol{s}}=\mathbf{1}$.  In case ({\bf B}) one has ${K_{01} \propto K_{10}^\dag \propto \sigma^{\pm}} \equiv (\sigma_x \pm i \sigma_y)/2$ while $K_{00},K_{11}$ both commute with $\sigma_z$.   
 Here $l_{\boldsymbol{s}}\neq \mathbf{1}$ and measurements at different times are not independent, but the post-selected trajectories simplify in a different way: for sufficiently large times the conditional state $ K^{\phantom{\dag}}_{i_{2t} j_{2t-1} } \!\!\cdots K^{\phantom{\dag}}_{i_1j_0 }|\psi\rangle$ must be an eigenstate of $\sigma^z$ and the system behaves classically. Transitions between the two eigenstates are signalled by $(1,0)$ and $(0,1)$ measurement pairs, which appear in an alternating sequence, interspersed by random $(0,0)$s and $(1,1)$s.

\begin{figure}[t!]
    \centering
    \includegraphics[width=1\columnwidth]{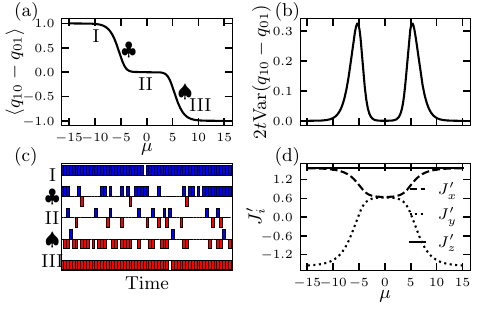}
    \caption{Illustration of the subcase ({\bf C}) for circuit parameters $J_x = 0.205 \pi$  and $J_y = 0.2\pi$, and tilting parameters $s_{00} = s_{11} = 0$ and $s_{01} = -s_{10}$. (a,b) Mean and (normalised) variance of the accumulated number of $(0,1)$ minus $(1,0)$ measurements. (c) Typical trajectories for $\mu = 0,\pm10$ [regimes (I,II,III)] and $\mu=\pm5.282$ [crossover points where $2t\text{Var}(q_{10} - q_{01})$ is maximised, marked by $(\clubsuit,\spadesuit)$]. Measurement outcomes $(1,0)$ and $(0,1)$ are marked in blue and red; $(0,0)$ and $(1,1)$ are not shown. (d) The new circuit parameters $\{ J'_{i} \}$ that realise the Doob transformation of the original circuit.
}
    \label{fig:figure3}
\end{figure}

The final case ({\bf C}) is not simple and exhibits an interesting dynamical crossover under post-selection.
We assume $J_x>J_y$ and $J_z=+\pi/2$ for concreteness.  Defining
$\mu \equiv (s_{10}-s_{01})/2$ and
 $\delta = \sin[(J_x -J_y)/2]/\cos[(J_x+J_y)/2]$, one finds
\be
\begin{aligned}
\tilde{K}_{10} &= A(\mu,\delta)
    \begin{pmatrix}
  0 &  \sqrt{\delta^2+e^{2\mu}}  \\ 
  -\delta \sqrt{1+\delta^2e^{2\mu}}   & 0
\end{pmatrix},\\ 
    \tilde{K}_{01} &= A(\mu,\delta)
    \begin{pmatrix}
  0 &  - \delta e^{\mu} \sqrt{\delta^2+e^{2\mu}}    \\ 
   e^{\mu} \sqrt{1+\delta^2e^{2\mu}}  & 0
\end{pmatrix},
\end{aligned}
\ee
while $\tilde{K}_{00}$ and $\tilde{K}_{11}$ only depend on $\mu$ through a multiplicative constant (see Appendix \ref{sec:example} for details). For $\delta \ll 1$ we have the following three regimes: 
\begin{align}
&{\rm (I)}& &\!\!\!\tilde{K}_{10} \propto \sigma_{y}, \quad \tilde{K}_{01} \propto  \delta^{-1} e^{\mu} \sigma_-, \quad e^{\mu}\ll \delta\,,\label{eq:KrausI}\\
&{\rm (II)}& &\!\!\!\tilde{K}_{10} \propto  \sigma_{+},\quad \tilde{K}_{01} \propto  \sigma_-, \quad  \delta \!\ll \!e^{\mu}\!\ll\! {\delta^{-1}},\label{eq:KrausII}\\ 
&{\rm (III)}& &\!\!\!\tilde{K}_{10} \propto \delta^{-1} e^{-\mu} \sigma_{+},\quad \tilde{K}_{01} \propto \sigma_y,\quad e^{\mu}\gg {\delta^{-1}}\,,\label{eq:KrausIII}
\end{align}
as shown in Figure~\ref{fig:figure3}. Regime (II) includes typical trajectories ($\mu\approx0$): it resembles case ({\bf B}) in that (1,0) and (0,1) alternate, interspersed with (0,0)s and (1,1)s, although corrections to \eqref{eq:KrausII} at $O(\delta)$ mean that this dynamics is still non-classical. In contrast, regimes (I,III) describe trajectories that are post-selected for an overwhelming majority of either (1,0) or (0,1) measurements, breaking the alternating structure. All three regimes have highly-structured measurement records (either alternating or dominated by one outcome) and are separated by crossovers where the records have much larger variance, Fig.~\ref{fig:figure3}(b,c) and the conditional state features strongly non-classical correlations.

The behaviour in the three regimes can be reproduced by a suitable $\tU$ gate, whose couplings are plotted in Fig.~\ref{fig:figure3}(d). From Eq.~\eqref{eq:channel} the eigenvalues of the channel are given in terms of the couplings by
$(1, \sin{J_x'} \sin{J_y'}, \sin{J_x'}\sin{J_z'}, \sin{J_y'} \sin{J'_z})$ (see Appendix \ref{sec:example}), meaning that the spectral gap of the Doob channel does not close during the crossovers. Instead, $\tU$ crosses over from $\tU \simeq U$ in regime (II) to an almost dual-unitary gate in regimes (I,III). The sharpness of this crossover reflects the rigid alternating structure of typical trajectories, so that large counting fields $|\mu|\sim \log(1/\delta)$ are required to post-select any other outcomes.

\section{Conclusion} 

We studied the interplay between unitary dynamics and rare measurements outcomes in quantum circuits. We showed that a class of post-selection problems, specifically that of biasing dynamics according to the large deviations in the number of measurement outcomes along the light-cone, can be 
resolved by mapping to a \emph{different} circuit for which the rare events are the (easy to access) typical behaviour. This provides an efficient way to access properties of quantum trajectories in real experiments avoiding post-selection overheads. 

Even though our results are obtained in a heavily simplified setting, they are in our view an important proof of principle: it is sometimes possible to embed the effect of measurements in a quantum many-body system by studying the purely unitary dynamics of a different system. The method that we have developed to achieve this goal is general and can be applied to more realistic scenarios. First, as we briefly discuss in Appendix~\ref{sec:general_unitary}, one can extend our treatment to cases where $U\neq U_{\rm{XYZ}}$. One may also apply it to describe the quantum trajectories for finite times going beyond the large time limit considered here~\cite{carollo2018making}. Another important application will be to consider truly many-body quantum channels that can reveal measurement-induced phase transitions. A promising arena where to study this question is that of quantum circuits with additional structure, such as integrability or dual-unitarity, which can give access to the fixed point of many-body channels~\cite{vanicat2018integrable, giudice2022temporal}.

\begin{acknowledgments}
  We acknowledge financial support from EPSRC Grant No.\ EP/V031201/1 (J.\ P.\ G.), from the Royal Society through the University Research Fellowship No.\ 201101 (B.\ B.). Support from Downing College is acknowledged (J.\ L.). 
\end{acknowledgments}

\footnotetext[10]{{\color{blue}See the Supplemental Material for: (a) A detailed explanation of the diagrammatic notation; (b) The explicit form of the light cone channel and of the Kraus operators for qubit systems; (c) A review of large deviation methods for quantum trajectories; (d) A simple formula relating the tilted (deformed) channel and the new unitaries; (e) Further details of the example discussed in the main text; (f) A comment on the effects of local unitaries.} 
}

\onecolumngrid
\begin{appendix}
\section{Diagrammatic Representation}
\label{sec:diagram}

Quantum circuits are conveniently described using a diagrammatic representation borrowed from tensor-network theory \cite{cirac2021matrix}. For a two-qubit unitary $U$, the matrix elements and the complex conjugate are represented by  
\begin{align}
    \mel{im}{U}{nj} &= 
    \text{
        \fineq[-0.8ex][1][1]{
            \tsfmatV[0][-0.5][r][1][][][bertinired]
            \node at (-0.6,-0.15) {$n$};
            \node at (0.6,-0.15) {$j$};
            \node at (-0.6,1.15) {$i$};
            \node at (0.6,1.15) {$m$};
        }
    } \quad 
    \text{and} \quad 
    \mel{im}{U^*}{nj} = 
    \text{
        \fineq[-0.8ex][1][1]{
            \tsfmatV[0][-0.5][r][1][][][bertiniblue]
            \node at (-0.6,-0.15) {$n$};
            \node at (0.6,-0.15) {$j$};
            \node at (-0.6,1.15) {$i$};
            \node at (0.6,1.15) {$m$};
        }
    }.
\end{align}

Using the representation above, many-body brickwork circuits $\mathbb{U}$ defined by Eq. (\ref{eq:U}) are built from the local two-qubit unitary $U$, for example, the diagrammatic representation for brickwork circuits of $2L=6$ qubits is 
\begin{align}
    \mathbb{U} = 
    \text{
        \fineq[-0.8ex][.65][1]{
            \foreach \i in {0,...,2}
            {
                \roundgate[2*\i][0][1][topright][bertinired][-1]
                \roundgate[2*\i+1][1][1][topright][bertinired][-1]
            }
            \hook[-1][0][b][r]
            \hook[-1][1][t][r]
            \hook[5][1][t][l]
            \hook[5][2][b][l]
            \hook[5][0][b][l]
            \hook[-1][2][b][r]
            \draw[gray, dotted] (-.5,0.25) -- (5.5,0.25);
            \draw[gray, dotted] (-.5,-0.25) -- (5.5,-0.25);
        }
    }
\end{align}
where we have used straight (connected) lines to represent summing over the indexes. Here and throughout, hooks at the boundaries of the many-body circuits shown represent contractions across the periodic boundaries. 
Similarly for the Hermitian conjugate:
\begin{align}
    \mathbb{U}^{\dagger} = 
    \text{
        \fineq[-0.8ex][.65][1]{
            \foreach \i in {0,...,2}
            {
                \roundgate[2*\i+1][0][1][bottomright][bertiniblue][-1]
                \roundgate[2*\i][1][1][bottomright][bertiniblue][-1]
            }
            \hook[-1][1][b][r]
            \hook[-1][2][t][r]
            \hook[5][0][t][l]
            \hook[5][1][b][l]
            \hook[5][2][t][l]
            \hook[-1][0][t][r]
            \draw[gray, dotted] (-.5,0.75) -- (5.5,0.75);
            \draw[gray, dotted] (-.5,1.25) -- (5.5,1.25);
        }
    }
\end{align}

The unitary condition $UU^{\dagger} = \mathbf{1} = U^{\dagger}U$  is represented by the following diagram
\begin{align}
\fineq[-0.8ex][1][1]{
\tsfmatV[0][1.25][r][1][][][bertinired][topright]
\tsfmatV[0][0][r][1][][][bertiniblue][bottomright]
\draw[ thick] (-0.5,1.5) -- (-0.5,1.75);
\draw[ thick] (0.5,1.5) -- (0.5,1.75);}
 =
\fineq[-0.8ex][.65][1]{
\draw[ thick] (-0.45,0.15) -- (-0.45,2.35);
\draw[ thick] (0.45,0.15) -- (0.45,2.35);
} 
= \fineq[-0.8ex][1][1]{
\tsfmatV[0][1.25][r][1][][][bertiniblue][bottomright]
\tsfmatV[0][0][r][1][][][bertinired][topright]
\draw[ thick] (-0.5,1.5) -- (-0.5,1.75);
\draw[ thick] (0.5,1.5) -- (0.5,1.75);}\,, 
\label{equ:unitary}
\end{align}
where the straight lines in the middle equation represent the identity. The diagrams provide a visual way of performing calculations and expressing physical quantities. The light-cone channel Eq. (\ref{eq:channel}) can be represented as 
\begin{align}
    {\cal M}_U [\rho] = \frac{1}{2} \quad
    \text{
        \fineq[-0.8ex][1][1]{
            \tsfmatV[0][1.25][r][1][][][bertinired][topright]
            \tsfmatV[0][0][r][1][][][bertiniblue][bottomright]
            \draw[thick] (-0.8,0.5) -- (-0.8,2.75);
            \draw[thick] (-0.8,2.75) -- (-0.5,2.75);
            \draw[thick] (-0.8,0.5) -- (-0.5,0.5);
            \draw[thick] (0.5,1.5) -- (0.5,1.75);
            \node at (-0.6, 1.6) {$\rho$};
        }
    }
\end{align}

The focus of the paper is the unravelling of light-cone channels into quantum trajectories. We introduce further diagrammatic representation. Fig. \ref{fig:TNdiagrams}(b) illustrates our measurement scheme, where measurements are indicated by red crosses. Fig. \ref{fig:TNdiagrams}(c) illustrates a quantum trajectory corresponding to a particular measurement record (a set of measurement outcomes, each in $\{0, 1\}$).  These measurements appear as red dots with their values indicated; mathematically, the red dots are defined in the main text as projectors $P^{(i)}=|i\rangle\langle i|$. 
For measurement outcome  $ij$, the 
action on the state is given by the single-qubit Kraus operator $K_{ij}$
\begin{align}
    K_{ij} = \frac{1}{\sqrt{2}} 
    \text{
        \fineq[-0.8ex][1][1]{
            \tsfmatV[0][-0.5][r][1][][][bertinired];
            \node at (0.6,-0.15) {$j$};
            \node at (-0.6,1.15) {$i$};
        }
    }.
\end{align}
(see also Eq.~(\ref{eq:Kloc}) in the main text).

To obtain Fig. \ref{fig:TNdiagrams}(d) we use the unitarity formula Eq. (\ref{equ:unitary}) repeatedly in Fig.~\ref{fig:TNdiagrams}(c) \cite{bertini2019exact}.

\section{Light cone channel}
\label{sec:light-cone-channel}
\subsection{Parameterisation of two-qubit gates}
To parameterise the brickwork circuits considered here, note that a general two-qubit unitary $U \in SU(4)$ enjoys the following decomposition known as the Cartan form \cite{zhang2003geometric}
\begin{equation}
     U =  \left( u_{3} \otimes u_{4} \right) U_{\text{XYZ}} \left( u_{1} \otimes u_{2} \right),
\label{eq:general_unitary}
\end{equation}
where $\{ u_{i} \} \in SU(2)$ and $U_{\text{XYZ}}$ is  the non-local unitary gate \cite{kraus2001optimal},
\begin{align}
    \begin{split}
        U_{\text{XYZ}} & = \exp[-\frac{i}{2} \biggl( J_x \sigma_x \otimes \sigma_x + J_y \sigma_{y} \otimes \sigma_{y} + J_z \sigma_z \otimes \sigma_z \biggr) ]  \\ 
        & = 
    \begin{pmatrix}
 e^{-iJ_z/2} \cos(\frac{Jx-Jy}{2}) & 0 & 0 &  -ie^{-iJ_z/2} \sin(\frac{Jx-Jy}{2}) \\ 
  0 & e^{iJ_z/2} \cos(\frac{Jx+Jy}{2}) & -ie^{iJ_z/2} \sin(\frac{Jx+Jy}{2}) & 0 \\
  0 & -ie^{iJ_z/2} \sin(\frac{Jx+Jy}{2}) & e^{iJ_z/2} \cos(\frac{Jx+Jy}{2}) & 0 \\
   -ie^{-iJ_z/2} \sin(\frac{Jx-Jy}{2}) & 0 & 0 &  e^{-iJ_z/2} \cos(\frac{Jx-Jy}{2})
\end{pmatrix}
    \end{split}
\end{align}
where $(J_x,J_y,J_z) \in [0,\pi]$ and $\sigma_{i}$ are the spin-$\frac{1}{2}$ Pauli matrices. In this work, we consider these non-local gates on the extended domain $(J_x,J_y,J_z) \in [-\pi, \pi]$, which we call XYZ unitaries here. 

The local unitary $u \in SU(2)$ (dropping the subscript) is parameterized as 
\begin{equation}
    u(\theta,\phi,\psi) = 
    \begin{pmatrix}
  \cos(\theta/2) e^{i\phi /2} & -\sin(\theta/2)e^{i\psi/2}\\ 
  \sin(\theta/2)e^{i\psi/2} & \cos(\theta/2) e^{-i\phi /2}
\end{pmatrix},
\label{eq:smallu-parametrisation}
\end{equation}
where $\theta \in [0,\pi]$, ($\phi, \psi) \in [0,4\pi]$. Arbitrary two-qubit unitaries in $U(4)$ are related to $U \in SU(4)$ by multiplying a complex phase which plays no role in this work.

\subsection{Parameterisation of the light cone channel}
The light cone channels introduced in the main text are a family of single-qubit channels. We focus on the light cone channels $\mathcal{M}_{U}[\cdot]$, which governs the two-point functions along the light cone in unitary circuits \cite{bertini2019exact}. For a given unitary circuit $U$, the light cone channel defines the state $\rho'$ at the next discrete time step
\begin{equation}
    \rho' = \mathcal{M}_{U}[\rho] = \frac{1}{2} \Tr_A \Big[ U (\rho \otimes \mathbf{1} _B) U^{\dagger} \Big],
\label{eq:light-cone-channel}
\end{equation}
where $\rho$ is defined on subsystem A and the subscripts $A$ and $B$ denote a subsystem its complement; also $\dagger$ is the conjugate transpose; the subscript $A$ is dropped in the following. Note that the trace acts on subsystem $A$, in contrast to `tracing the environment out' in standard considerations of quantum channels. Another obvious but worth-mentioning fact is that the subsystem $B$ is the maximally mixed state rather than a simple pure state. 

For any unitary, the light cone super-operator $| \rho ' \rangle= \mathcal{M}_{U} | \rho \rangle $ has the following decomposition
\begin{equation}
    \mathcal{M}_{U} = (u_4 \otimes u_4^{*}) \mathcal{M}_{U_{\text{XYZ}}} ( u_1 \otimes u_1^{*} ).
\label{eq:general_U_channel}
\end{equation}
From now on we drop the subscript $U_{\text{XYZ}}$ in $\mathcal{M}_{U_{\text{XYZ}}}$ unless otherwise stated.

In particular, this work focuses on the channel of the XYZ unitaries so $\mathcal{M}(J_x,J_y,J_z)$ take three real parameters and has the following explicit form
\begin{multline}
    \mathcal{M}(J_x,J_y,J_z)  =  \\ 
    \frac{1}{2} 
    \begin{pmatrix}
 1 + \sin(J_x) \sin(J_y)  & 0 & 0 &  1 - \sin(J_x) \sin(J_y) \\ 
  0 & (\sin(J_x) + \sin(J_y) ) \sin(J_z) & (-\sin(J_x) + \sin(J_y) ) \sin(J_z) & 0 \\
  0 & (-\sin(J_x) + \sin(J_y) ) \sin(J_z) & (\sin(J_x) + \sin(J_y) ) \sin(J_z) & 0 \\
  1 - \sin(J_x) \sin(J_y) & 0 & 0 &  1 + \sin(J_x) \sin(J_y)
\end{pmatrix},
\label{eq:Heisenberg-channel}
\end{multline}
where $(J_x,J_y,J_z) \in [-\pi,\pi]$. 

Furthermore, $\mathcal{M}(J_x,J_y,J_z)$ enjoys the following special properties that enable further analytical insights. The hermiticity $\mathcal{M}^{\dagger}(J_x,J_y,J_z) = \mathcal{M}(J_x,J_y,J_z)$ guarantee a real spectrum. The four eigenvalues $\{ \lambda_i \}$ are $ \lambda_0 = 1$, $\lambda_1 = \sin(J_x)\sin(J_y)$, $\lambda_2= \sin(J_x)\sin(J_z)$ and $\lambda_3 = \sin(J_y)\sin(J_z) $. 

The spectrum has the following structure. The real values of $\{ J_{i} \}$ introduce a set of constraints on the allowed spectrum and they satisfy the following set of inequalities
\begin{align}
\begin{split}
\lambda_1 & \ge  \lambda_2 \lambda_3 \\
\lambda_2 & \ge  \lambda_1 \lambda_3 \\
\lambda_3 & \ge  \lambda_1 \lambda_2. \\
\end{split}
\end{align}  

Moreover, the channels are essentially defined by terms in a standard XYZ Hamiltonian, and also inherit the $\mathbb{Z}_2$ symmetries $(\sigma_x \otimes \sigma_x)\mathcal{M}(J_x,J_y,J_z)(\sigma_x \otimes \sigma_x) = \mathcal{M}(J_x,J_y,J_z)$, or equivalently ${\cal M}_{U_{\rm XYZ}}(\sigma_x\rho\sigma_x)=\sigma_x {\cal M}_{U_{\rm XYZ}}(\rho)\sigma_x$. The same symmetry holds also for $\sigma_{y}$ and $\sigma_{z}$.

\section{Kraus Form}
\label{sec:unravelling}
\subsection{Environmental unravelling}
As discussed in the main text, we unravel the light cone channel in the computational basis.
The Kraus operators are defined as matrix elements of the light cone super-operator. For a unitary $U$ with matrix elements 
$(U)_{n j}^{i m} = \langle i m | U | n j  \rangle $, we define the Kraus operators to be
\begin{align}
\begin{split}
(K_{00})_{mn} = \frac{1}{\sqrt{2}} \langle 0 m | U  |n 0  \rangle \\
(K_{01})_{mn} = \frac{1}{\sqrt{2}} \langle 0 m| U  |n 1 \rangle \\
(K_{10})_{mn} = \frac{1}{\sqrt{2}} \langle 1 m| U |n 0 \rangle \\
(K_{11})_{mn} = \frac{1}{\sqrt{2}} \langle 1 m | U  | n 1\rangle
\end{split}
\label{eq:Computational-Kraus}
\end{align}
which corresponds to performing measurements along the light cone with four possible outcomes $\{ 00,01,10,11\}$. This definition follows naturally from the matrix elements of the light cone channel Eq. (\ref{eq:light-cone-channel}) in the computational basis
\begin{equation}
   \langle i | \Tr_A \Big[ U (\rho \otimes \mathbf{1} _B) U^{\dagger} \Big] |j \rangle =  \sum_k \langle k i | U (\rho \otimes \mathbf{1} _B) U^{\dagger}  |k j \rangle = \sum_{k,m,n,q} \langle k i | U | nq \rangle  \langle n| \rho | m\rangle  \langle mq| U^{\dagger}  |k j \rangle.
\end{equation}

For the XYZ unitaries $U_{\text{XYZ}}(J_x, J_y,J_z)$, the above definition gives the following explicit form of Kraus operators
\begin{equation}
    K_{00} = \frac{1}{\sqrt{2}}
    \begin{pmatrix}
  e^{-\frac{i J_z}{2}} \cos(\frac{J_x - J_y}{2}) & 0  \\ 
   0 & -i e^{\frac{i J_z}{2}} \sin(\frac{J_x + J_y}{2})
\end{pmatrix}
\quad
    K_{11} = \frac{1}{\sqrt{2}}
\begin{pmatrix}
  -i e^{\frac{i J_z}{2}} \sin(\frac{J_x + J_y}{2}) & 0  \\ 
   0 & e^{-\frac{i J_z}{2}} \cos(\frac{J_x - J_y}{2})
\end{pmatrix}
\end{equation}

\begin{equation}
    K_{10} = \frac{1}{\sqrt{2}}
    \begin{pmatrix}
  0 & e^{\frac{i J_z}{2}} \cos(\frac{J_x + J_y}{2})  \\ 
   -ie^{-\frac{i J_z}{2}} \sin(\frac{J_x - J_y}{2}) & 0
\end{pmatrix}
\quad
    K_{01} = \frac{1}{\sqrt{2}}
    \begin{pmatrix}
  0 &  -ie^{-\frac{i J_z}{2}} \sin(\frac{J_x - J_y}{2}) \\ 
   e^{\frac{i J_z}{2}} \cos(\frac{J_x + J_y}{2}) & 0
\end{pmatrix}.
\label{eq:Environmental-Kraus-01}
\end{equation}
As usual, the Kraus decomposition is related to the superoperator of XYZ channel Eq. (\ref{eq:Heisenberg-channel}) via $\mathcal{M}(J_x,J_y,J_z) = \sum_{ij} K_{ij} \otimes K^{*}_{ij}$.

A similar light cone channel can be defined for any two-qubit unitary $U$. The Kraus operators for a general unitary given by Eq. (\ref{eq:general_unitary}) are given by the corresponding XYZ Kraus operators $\{ K_{ij} \}$ multiplied by local unitaries as $\{ u_4 K_{ij}u_{1} \}$.

\section{Large deviations}
\label{sec:largde_deviation}

In this section, we show how to apply large deviation techniques to study the counting statistics of measurement outcomes of a quantum trajectory \cite{touchette2018introduction,garrahan2018aspects,jack2020ergodicity}.  We derive conditions under which the Doob channel is unital.

Consider a quantum trajectory of length $T=2t$, corresponding to a sequence of measurement outcomes $(\boldmath{i},\boldmath{j})$ in the main text. Let $K_{ij}$ be the number of times that measurement $(\boldmath{i},\boldmath{j})$ appears in the trajectory.  We collect these accumulated counts in a vector $\mathbf{Q}=(K_{00},K_{10},K_{01},K_{11})$.  The probability $\text{Prob}(\bold{Q})$ of observing a certain $\bold{Q}$ satisfies the large deviation principle in the large $t$ limit, 
\begin{equation}
    \text{Prob}(\bold{Q}) \simeq e^{-T F( \bold{q} )},
\end{equation}
where $\bold{q} = \bold{Q}/T$ and $F(\bold{q})$ is the rate function. The rate function determines the typical measurement outcomes given by the minimizer, and fluctuations are exponentially suppressed. 

Rare measurement outcomes can be made typical by defining a tilted (deformed) ensemble of trajectories, known as the s-ensemble, with the probability of finding the desired outcomes 
\begin{equation}
    \text{Prob}(\bold{Q} | \bold{s}) =\frac{e^{-\bold{Q}\cdot\bold{s}} \text{Prob}(\bold{Q})}{Z(\bold{s})}, 
\end{equation}
where $\bold{s} = (s_{0},s_{1},...,s_{r-1})$ is the counting field and $Z(\bold{s}) = \sum_{\bold{Q}} e^{-\bold{Q}\cdot\bold{s}} \text{Prob}(\bold{Q})$ is the moment generating function and also shows the large deviation principle. 
\begin{equation}
    Z(\bold{s}) \simeq e^{T \theta(\bold{s})},
\end{equation}
where $\theta(\bold{s})$ is the scaled cumulant generating function (SCGF). 

The SCGF is related to the rate function by a Legendre transformation
\begin{equation}
    \theta(\bold{s}) = -\min_{\bold{q}}[\bold{q}\cdot\bold{s} + F(\bold{q})],
\end{equation}
and can be obtained by tilting the original quantum channel 
\begin{equation}
    e^{\theta(\bold{s})} = \max( \text{Spec} ( \mathcal{M}_{\bold{s}} ) ),
\end{equation}
where $\text{Spec} ( \cdot)$ denotes the spectrum and
\begin{equation}
    \mathcal{M}_{\bold{s}}[\cdot] = \sum_{ij} e^{-s_{ij}} K_{ij} \cdot K^{\dagger}_{ij}
\end{equation}
is the tilted channel. 

The tilted channel $\mathcal{M}_s$ is not trace-preserving. But a trace-preserving channel can be recovered by via Doob transformation by defining a new channel $  \tilde{\mathcal{M}}_{\bold{s}}[\cdot]$
\begin{align}
\begin{split}
    \tilde{\mathcal{M}}_{\bold{s}}[\cdot] & = \sum_{ij} \tilde{K}_{ij} \cdot \tilde{K}_{ij}^{\dagger} \\
    \tilde{K}_{ij} & = \frac{e^{-s_{ij}/2}}{e^{\theta(\bold{s})/2} } l^{1/2}_{\bold{s}} K_{ij} l^{-1/2}_{\bold{s}},
\end{split}
\label{eq:Doob-Kraus}
\end{align}
where $l_{\bold{s}}$ is the leading left eigenmatrix of $\mathcal{M}_{s}$ such that 
\begin{equation}
    \mathcal{M}^{\dagger}_{s}[l_{\bold{s}}] = e^{\theta(\bold{s})} l_{\bold{s}}
\end{equation}

\subsection{Unital constraint} 
\label{subsec:unital}
Here we show there is a simple constraint on the leading eigenmatrices for the Doob transformed channel to be unital. The Doob channel is trace-preserving $\tilde{\mathcal{M}}^{\dagger}_{\bold{s}}[\mathbf{1}] = \mathbf{1}$ by definition and unital if $\tilde{\mathcal{M}}_{\bold{s}}[\mathbf{1}] = \mathbf{1}$, which implies 
\begin{equation}
     \tilde{\mathcal{M}}_{\bold{s}}[\mathbf{1}] = e^{-\theta(\bold{s})}
     l^{1/2}_{\bold{s}} \mathcal{M}_{\bold{s}}[l^{-1}_{\bold{s}}] l^{1/2}_{\bold{s}} = \mathbf{1},
\end{equation}
equivalently, $\mathcal{M}_{s}[l^{-1}_{\bold{s}}]  = e^{\theta(\bold{s})} l^{-1}_{\bold{s}}$. This condition is the same as the definition of the right leading eigenmatrix of the tilted channel $\mathcal{M}_{s}[r_{\bold{s}}]  = e^{\theta(\bold{s})} r_{\bold{s}}$. Thus, unital Doob channels have their right and left leading eigenmatrices with the following property 
\begin{equation}
    l_{\bold{s}} r_{\bold{s}} = \mathbf{1}.
\end{equation}

\section{Doob transformed circuits for XYZ unitaries }
\label{sec:Doob_channel}
\subsection{Tilted XYZ Channel}
In this section, we analyze the structure of the tilted channel for XYZ unitaries. For a general counting field $\bold{s} = (s_{00},s_{10},s_{01},s_{11})$ conjugate to the measurement outcomes $\{00,10,01,11\}$, the super-operator $\mathcal{M}_{U,\bold{s}}(J_x,J_y,J_z,s_{00},s_{10},s_{01},s_{11})$ reads
\begin{equation}
    \mathcal{M}_{U,\bold{s}}(J_x,J_y,J_z,s_{00},s_{10},s_{01},s_{11}) = \sum_{i} e^{-s_{ij}} K_{ij} \otimes K^{*}_{ij} = \frac{1}{2}  
    \begin{pmatrix}
 m_{00} & 0 & 0 &  m_{03} \\ 
  0 & m_{11} & m_{12} & 0 \\
  0 & m_{12}^{*} & m_{11}^{*} & 0 \\
  m_{30} & 0 & 0 & m_{33}
\end{pmatrix}
\label{eq:tilted-channel}
\end{equation}
\begin{align*} 
m_{00} &= e^{-s_{00}} \cos^2(\frac{J_x - J_y}{2}) + e^{-s_{11}}\sin^{2}(\frac{J_x + J_y}{2})\\
m_{03} &= e^{-s_{10}}\cos^2(\frac{J_x + J_y}{2}) + e^{-s_{01}}\sin^{2}(\frac{J_x - J_y}{2})\\
m_{11} &= \frac{1}{2}i(-e^{iJ_z -s_{11}} + e^{-iJ_z - s_{00}})(\sin(J_x) + \sin(J_y))\\
m_{12} &= \frac{1}{2}i(e^{iJ_z - s_{10}} - e^{-iJ_z - s_{01}})(\sin(J_x) - \sin(J_y))\\
m_{30} &= e^{-s_{01}}\cos^2(\frac{J_x + J_y}{2}) + e^{-s_{10}}\sin^{2}(\frac{J_x - J_y}{2})\\
m_{33} &= e^{-s_{11}}\cos^2(\frac{J_x - J_y}{2}) + e^{-s_{00}}\sin^{2}(\frac{J_x + J_y}{2})
\end{align*}
Note that the block structure of the original channel $\mathcal{M}(J_x,J_y,J_z)$ is preserved under an arbitrary deformation. 

Moreover, the block structure not only provides a simple expression for the spectrum but also constrains the leading eigenvectors. The leading eigenmatrix of $\mathcal{M}(J_x,J_y,J_z)$comes either from the outer or the inner block, due to the direct sum structure. In the computational basis, this would mean the leading eigenmatrix $r_{\bold{s}}$ is purely diagonal or off-diagonal, as shown below
 \begin{equation}
     \{ \begin{pmatrix}
  r_{00} & 0  \\ 
  0 & r_{11}
\end{pmatrix} ,\quad     \begin{pmatrix}
  0 & r_{01}  \\ 
  r^{*}_{01} & 0
\end{pmatrix} \}.
\end{equation}
The leading eigenmatrix must be positive due to the CP condition of the channel, and $r_{\bold{s}}$ is only positive for the diagonal one. This shows that the leading eigenmatrix is defined by the outer block and the leading eigenvalue $e^{\theta(\bold{s})}$ is 
\begin{equation}
    e^{\theta(\bold{s})} = \frac{1}{4} \left( m_{00} + m_{33} + \sqrt{ (m_{00} - m_{33})^2 + 4m_{03}m_{30} } \right).
\end{equation}

\subsection{Doob Transformation}
The unital property of the channel is not necessarily preserved under both deformation and Doob transformation. As discussed in the main text, we impose the unital property of the Doob transformed channel here. In this section, we show all the possible deformations of XYZ unitaries that remain XYZ unitaries under Doob transformation. Namely, we find special sets of $\{ J_i\}$ and $\bold{s}$ such that the following equation hold
\begin{equation}
    \tilde{\mathcal{M}}_{U,\bold{s}}(J_x,J_y,J_z,s_{00},s_{10},s_{01},s_{11}) = \mathcal{M}_{\tilde{U}}(J_x', J_y' , J_z'),
\label{eq:Doob-Heisenberg}
\end{equation}
where $\{ J'_i  \}  \neq \{J_i \} $ in general. In the rest of the section, the subscript $U$ and $\tilde{U}$ are dropped.

The leading right and left eigenmatrices of the tilted XYZ channel $\bold{s} = (s_{00},s_{10},s_{01},s_{11})$ are
\begin{equation}
 r_{\bold{s}} = \begin{pmatrix}
 \frac{m_{00} - m_{33} +\sqrt{ (m_{00} - m_{33})^2 + 4m_{03}m_{30} }}{2m_{30}} & 0  \\ 
  0 & 1
\end{pmatrix}
\end{equation}
and 
\begin{equation}
    l_{\bold{s}} = \begin{pmatrix}
 \frac{m_{00} - m_{33} +\sqrt{ (m_{00} - m_{33})^2 + 4m_{03}m_{30} }}{2m_{03}} & 0  \\ 
  0 & 1
\end{pmatrix}.
\label{equ:l-matrix}
\end{equation}
We assume $m_{30},m_{03} >0$ to avoid singular behavior in the eigenmatrices, which is essential for the Doob transformation to exist. In fact, $m_{30},m_{03}$ can be zero for the case of $J_x = J_y = \pm\pi/2$, and we leave this special case for the end of the section.     

The product of the two leading eigenmatrices is
\begin{equation}
    l_{\bold{s}} r_{\bold{s}} = \begin{pmatrix}
 \frac{ \big( m_{00} - m_{33} +\sqrt{ (m_{00} - m_{33})^2 + 4m_{03}m_{30} } \big)^2}{4 m_{03} m_{30}} & 0  \\ 
  0 & 1
\end{pmatrix},
\end{equation}
which to be solved for the unital Doob unitaries constraint  $ l_{\bold{s}} r_{\bold{s}} \propto \mathbf{1}$ shown in section \ref{subsec:unital}. Taking $m_{00} = m_{33}$ and using the $m_{30},m_{03} >0$ assumption above, the Doob channel is unital if and only if 
\begin{equation}
    m_{33}  = m_{00}.
\end{equation}

This condition is always satisfied in the following two cases without imposing any conditions on the counting field 
\begin{itemize}
    \item Either $(J_x, J_z) = (\pm \pi/2, \pm \pi/2)$, or $(J_y, J_z) = (\pm \pi/2, \pm \pi/2)$ or $(J_x, J_y) = \pm ( \pi/2, - \pi/2)$; These correspond to dual-unitary gates $U$. The spectrum of ${\cal M}$ has two eigenvalues with $|\lambda|=1$, so one sees that these channels are not ergodic.  The existence of a quantum Doob transformation is not always guaranteed for non-ergodic channels~\cite{chetrite2015nonequilibrium}, but there are no difficulties with the examples considered here. Physically, this reflects that while quantum trajectories of these non-ergodic channels retain information about their initial condition for all times, the statistics of $\bold{Q}$ are independent of the initial condition.
\end{itemize}

We discuss some illustrative cases.
If $J_x=J_z=\pm\pi/2$ then  $l_{\bold{s}} = \mathbf{1} + c \sigma_y$ is a left eigenmatrix of ${\cal M}_{U,\bold{s}}$, independent of $c$.  This $l_{\bold{s}}$ commutes with all Kraus operators so there is a unique Doob transformation: $\tilde{K}_{ij}=e^{-(\theta(\bold{s})+s_{ij})/2}K_{ij}$. (We take $c\in(-1,1)$ so that $l_{\bold s}$ is CP, as required for any quantum Doob transform.) For the case $J_x=-J_z=\pm\pi/2$, there are eigenmatrices $l_{\bold s}=\mathbf{1}$ with eigenvalue $\theta(\bold{s})$ and $l^-_{\bold s}=\sigma_y$ with eigenvalue $-\theta(\bold{s})$.  Again, both eigenmatrices commute with all Kraus operators and the Doob transformation is simple: $\tilde{K}_{ij}=e^{-(\theta(\bold{s})+s_{ij})/2}K_{ij}$. The cases $(J_y,J_z) =(\pm \pi/2,\pm\pi/2)$ are exactly analogous, with $\sigma_y\to\sigma_x$. The cases $(J_x,J_y)=\pm(\pi/2,-\pi/2)$ are very similar. 

To explore solutions for a more generic choice of unitaries, two components of the counting field have to be
\begin{equation}
    s_{00} = s_{11},
\end{equation}
which leads to $m_{00} = m_{33}$ and $m_{11} = \sin(J_z)e^{-s_{00}}(\sin(J_x) + \sin(J_y))$ is real. 
The super-operator of the Doob transformed channel reads
\begin{equation}
    \tilde{\mathcal{M}}_{\bold{s}}(J_x,J_y,J_z,s_{00},s_{10},s_{01},s_{00}) = \frac{1}{e^{\theta(s)}} \begin{pmatrix}
 m_{00} & 0 & 0 & \sqrt{m_{30}} \sqrt{m_{03}} \\ 
  0 & m_{11} & m_{12} & 0 \\
  0 & m_{12}^{*} & m_{11} & 0 \\
  \sqrt{m_{30}} \sqrt{m_{03}} & 0 & 0 & m_{00}
\end{pmatrix}, 
\end{equation}
and to solve Eq. (\ref{eq:Doob-Heisenberg}), the channel $\tilde{\mathcal{M}}_{\bold{s}}$ must have $\mathbb{Z}_2$ symmetry. We check this by writing  
\begin{equation}
    (\sigma_x \otimes \sigma_x )\tilde{\mathcal{M}}_{\bold{s}}(J_x,J_y,J_z,s_{00},s_{10},s_{01},s_{00})(\sigma_x \otimes \sigma_x ) - \tilde{\mathcal{M}}_{\bold{s}}(J_x,J_y,J_z,s_{00},s_{10},s_{01},s_{00})= 
    \frac{1}{e^{\theta(s)}} \begin{pmatrix}
 0 & 0 & 0 & 0 \\ 
  0 & 0 & p & 0 \\
  0 & p & 0 & 0 \\
  0 & 0 & 0 & 0
\end{pmatrix},
\end{equation}
where
\begin{align*} 
p &= i \cos(J_z)e^{-s_{10}-s_{01}}(-e^{-s_{10}} + e^{-s_{01}})(\sin(J_x) - \sin(J_y))\\
\end{align*}
So preserving $\mathbb{Z}_2$ symmetry requires $p=0$ which can be achieved in three different ways:
\begin{itemize}
    \item $s_{10} = s_{01}$; The counting field for $K_{01}$ and $K_{10}$ have to be same and $\{ J_i \}$ are three free parameters. The leading eigenmatrix is $\mathbf{1}$.   
    \item $\sin(J_x) = \sin(J_y)$; Two of the Kraus operators have the following simple form, $K_{10} \propto \sigma^{+}$ and $K_{01} \propto \sigma^{-}$. The leading left eigenmatrix is non-trivial: \begin{equation}
    l_{\bold{s}} = \begin{pmatrix}
  e^{(s_{10} -s_{01})/2} & 0  \\ 
  0 & 1
\end{pmatrix}
\end{equation}.
    \item $\cos(J_z) = 0$; This case is discussed in the main text extensively. The leading left eigenmatrix is non-trivial, 
    \begin{equation}
    l_{\bold{s}} = \begin{pmatrix}
  \sqrt{\frac{e^{-s_{01}}\cos^{2}(\frac{J_x+J_y}{2}) + e^{-s_{10}}\sin^{2}(\frac{J_x-J_y}{2}) }{e^{-s_{10}}\cos^{2}(\frac{J_x+J_y}{2}) + e^{-s_{01}}\sin^{2}(\frac{J_x-J_y}{2})}} & 0  \\ 
  0 & 1
\end{pmatrix}.
\end{equation}
\end{itemize}

Now we come back to the special case mentioned above,
\begin{itemize}
    \item For $J_x = J_y= \pm \pi/2$ the circuits are dual unitaries.  One has $m_{30} = m_{03} = 0$ and $m_{00} = m_{33}$ so Eq. \eqref{equ:l-matrix} is not applicable but it is easily seen from Eq. (\ref{eq:tilted-channel}) that  $l_{\bold{s}} = \mathbf{1} + c \sigma_z$ is an eigenmatrix of ${\cal M}_{U,\bold{s}}$ for all $c$. As in the previous dual unitary cases, this $l_{\bold{s}}$ commutes with all Kraus operators so the quantum Doob transformation exists with $\tilde{K}_{ij}=e^{-(\theta(\bold{s})+s_{ij})/2}K_{ij}$.  For these parameters,  there are only two Kraus operators in the environmental unravelling, $K_{01}=0=K_{10}$ in Eq. (\ref{eq:Environmental-Kraus-01}).
\end{itemize}

For all the above choices of parameters, we can solve for $\mathcal{M}(J'_x, J'_y,J'_z)$. First, we introduce a compact notation by noting that the matrix elements of the tilted channel $\mathcal{M}_{s}$ are closely related to its spectrum $\{ \lambda(\bold{s})_{i} \}$: 
\begin{align}
\begin{split}
e^{\theta(\bold{s})}=\lambda_0(\bold{s}) &= m_{00} + \sqrt{m_{30}m_{03}} \\
\lambda_1(\bold{s}) &= m_{00} - \sqrt{m_{30}m_{03}} \\
\lambda_2(\bold{s}) &= m_{11} + m_{12}\\
\lambda_3(\bold{s}) &= m_{11} - m_{12}.
\end{split}
\end{align}

The set of $\{ J'_{i} \}$ is given by solving the following two sets of equations: 
\begin{align}
\begin{split}
\sin^{2}(J_x') &=  \frac{ \lambda_3(\bold{s}) \lambda_1(\bold{s})}{\lambda_2(\bold{s}) \lambda_0(\bold{s})} \quad \quad \quad \frac{\sin(J_x')}{\sin(J_y')} =  \frac{ \lambda_3(\bold{s}) }{\lambda_2(\bold{s}) }\\
\sin^{2}(J_y') &=  \frac{ \lambda_2(\bold{s}) \lambda_1(\bold{s})}{\lambda_3(\bold{s}) \lambda_0(\bold{s})} \quad \quad \quad \frac{\sin(J_z')}{\sin(J_x')} =  \frac{ \lambda_2(\bold{s}) }{ \lambda_1(\bold{s})}\\
\sin^{2}(J_z') &=  \frac{ \lambda_2(\bold{s}) \lambda_3(\bold{s})}{\lambda_1(\bold{s}) \lambda_0(\bold{s})} \quad \quad \quad \frac{\sin(J_z')}{\sin(J_y')} =  \frac{ \lambda_3(\bold{s})}{\lambda_1(\bold{s})}.
\end{split}
\label{eq:new_J}
\end{align}

Note that an arbitrary choice of parameters may exceed the range of $\sin(J_{i})$, and can lead to complex-valued $\{ J'_i \}$. Such complex-valued $\{ J'_i \}$ reproduces the correct Doob transformed channel in Eq. (\ref{eq:Doob-Heisenberg}), but the circuits are no longer unitary and will be excluded here.

Overall, there are four classes of parameters for which Eq. (\ref{eq:Doob-Heisenberg}) can be solved, so that $\tilde{\mathcal{M}}_{s}$ can be obtained as a trace over a brickwork unitary circuit
\begin{itemize}
  \item Any two of $\{J_i \} = \pm \pi/2$  
  \item $s_{00} = s_{11}$ and $s_{10} = s_{01}$ 
  \item $s_{00} = s_{11}$ and $\sin(J_x) = \sin(J_y)$
  \item $s_{00} = s_{11}$ and $\cos(J_z) = 0$.
\end{itemize}

\subsection{Making rare measurement outcomes typical}
In this section, we show how to build the equality between the rare measurement outcomes of $\{ J_i \}$ and typical ones in $\{ J'_i \}$ at the level of quantum trajectories as well as the level of accumulated counting statistics. To see this concretely, consider a trace-preserving set of Kraus operators and normalized initial pure state $ |\psi_0 \rangle$. The underlying stochastic process of the completely positive trace-preserving map is given by 
\begin{equation}
  |\psi_{t+1} \rangle \langle \psi_{t+1} | = 
 \frac{ K_{ij} |\psi_t \rangle \langle \psi_t | K^{\dagger}_{ij} }{\Tr[K_{ij} |\psi_t \rangle \langle \psi_t | K^{\dagger}_{ij}]} 
\label{eq:update-prob}
\end{equation}
with probability $\Tr[K_{ij} |\psi_t \rangle \langle \psi_t | K^{\dagger}_{ij}]$.

Therefore, for a normalised initial state $\Tr(|\psi_0 \rangle \langle \psi_0 |) = 1$, the probability of observing a tilted trajectory of length $T = 2t$ is given by
\begin{align}
\begin{split}
     \text{Prob}[i_{1}j _{0} \cdots i_{2t}j_{2t-1} | \bold{s}] & = \frac{1}{Z(\bold{s})}\Tr[ K_{i_{2t} j_{2t-1}}e^{-s_{i_{2t}j_{2t-1}}/2} \cdots K_{i_1 j_0} e^{-s_{i_1 j_0}/2}|\psi_0 \rangle \langle \psi_0 | e^{-s_{i_1 j_0}/2} K_{i_1 j_0}^{\dagger} \cdots e^{-s_{i_{2t}j_{2t-1}}/2}K_{i_{2t} j_{2t-1}}^{\dagger} ] \\ 
     & =  \frac{e^{2t\theta(\bold{s})}}{Z(\bold{s})}\Tr[  l^{1/2}_{\bold{s}} \tilde{K}_{i_{2t} j_{2t-1}} \cdots \tilde{K}_{i_1 j_0} l^{-1/2}_{\bold{s}} |\psi_0 \rangle \langle \psi_0 |l^{-1/2}_{\bold{s}} \tilde{K}_{i_1 j_0}^{\dagger} \cdots \tilde{K}_{i_{2t} j_{2t-1}}^{\dagger}  l^{1/2}_{\bold{s}}].
\end{split}
\end{align}
Note that the summand in Eq. (\ref{eq:biasedsum}) of the main text is realised by taking $b = |\psi_0 \rangle \langle \psi_0 |$ and $a = \mathbf{1}$.

In the large-deviation limit $e^{2t\theta(\bold{s})}/Z(\bold{s}) = O(1)$, the probability is the same as the Doob transformed trajectory up to a transient difference close to the first and the last time steps
\begin{align}
\begin{split}
    \text{Prob}[i_{1}j _{0} \cdots i_{2t}j_{2t-1} | \bold{s}] & \simeq \Tr[ \tilde{K}_{i_{2t} j_{2t-1}} \cdots \tilde{K}_{i_1 j_0}|\psi_0 \rangle \langle \psi_0 |\tilde{K}_{i_1 j_0 }^{\dagger} \cdots \tilde{K}_{i_{2t} j_{2t-1} }^{ \dagger} ] . 
\end{split}
\end{align}
The accumulated counting statistics follow directly from the above 
\begin{equation}
    \langle \bold{Q} \rangle_{\bold{s}}(J_x,J_y,J_z) =  \langle \tilde{ \bold{Q}} \rangle(J_x,J_y,J_z),
\end{equation}
as shown in Ref. \cite{carollo2018making}. 

As discussed in the main text, we aim to reproduce the Doob transformed channel as the light cone channel of another XYZ circuit $\tilde{U}$ with parameters $(J'_x,J'_y,J'_z)$
\begin{equation}
    \tilde{\mathcal{M}}_{\bold{s}}[\rho] = \frac{1}{2} \Tr_A \Big[ \tilde{U} (\rho \otimes \mathbf{1} _B) \tilde{U}^{\dagger} \Big].
\end{equation}
The relevant superoperator is given in Eq. (\ref{eq:Doob-Heisenberg}) and the Kraus operators $\{ K'_{ij} \}$ in the computational basis are given by Eq. (\ref{eq:Computational-Kraus}). Note that the quantum trajectories are not the same 
\begin{equation}
    \text{Prob}[i_{1}j _{0} \cdots i_{2t}j_{2t-1} | \bold{s}] \neq \Tr[ K'_{i_{2t} j_{2t-1} } \cdots K'_{i_1 j_0 } |\psi_0 \rangle \langle \psi_0 | K_{i_1 j_0 }^{'\dagger} \cdots K_{i_{2t} j_{2t-1} }^{'\dagger} ],
\end{equation}
despite that $\{ K'_{ij} \}$ and $\{ \tilde{K}_{ij} \}$ produce the same channel. 

As a unique feature of quantum mechanics without classical (stochastic process) analog, the same quantum channel can be decomposed into different sets of Kraus operators, $\{ K^{\text{set1}}_{ij} \}$ and $\{ K^{\text{set2}}_{ij} \}$ that are related by an isometry $V$,
\begin{equation}
    K^{\text{set1}}_{ij} = \sum_{i' j'}V_{ij,i'j'}K^{\text{set2}}_{i'j'}.
\label{eq:Kraus-change-basis}
\end{equation}
For example, the Doob transformed Kraus operator $\{ \tilde{K}_{ij} \}$ and the environmental unravelling $\{K'_{ij} \}$ of new unitaries for $(J'_x,J'_y,J'_z)$ generate the same channel. However, they produce different quantum trajectories.

As indicated in the above equation, the probability of observing a certain trajectory depends on the choice of Kraus operators even for the same quantum channel. Therefore, one has to transform $\{ K'_{ij} \}$ into the Doob transformed Kraus  $\{ \tilde{K}_{ij} \}$ to achieve the rare-typical mapping at the single-trajectory level. Such basis transformation can be computed directly since the explicit forms of two sets of Kraus are known. Let $\tilde{\bold{e}}$ be the basis formed by the basis vector $| \tilde{K}_{ij} \rangle =  (\tilde{K}_{ij})_{mn} |m n \rangle$, 
\begin{equation}
    \tilde{\bold{e}} =
\begin{bmatrix}
    \vert & \vert & \vert & \vert\\
    | \tilde{K}_{00} \rangle &| \tilde{K}_{10} \rangle & | \tilde{K}_{01} \rangle & | \tilde{K}_{11} \rangle   \\
    \vert & \vert & \vert & \vert
\end{bmatrix},
\end{equation}
and similarly for $\bold{e}'$ being the basis formed by $| K'_{i'j'} \rangle =  (K'_{i'j'})_{m'n'} |m' n' \rangle =  (\tilde{U}^{\dagger})_{n' j'}^{ i' m'}|m' n' \rangle / \sqrt{2}$. Note that the basis is non-orthogonal in general.

Then, the tilted probability is recovered by acting $V = \left( \bold{e}^{' -1} \tilde{\bold{e}} \right)^{T}$ on the computational basis of the new unitary, such that  
\begin{equation}
    \tilde{K}_{ij} = \sum_{i'j'} V_{ij,i'j'} K'_{i'j'}.
\end{equation}

Here we find the basis transformation for the XYZ unitaries $U_{\text{XYZ}}$ explicitly. Due to the block structure, the Kraus operators for $\{ 00, 11\}$ are decoupled from $\{ 10, 01\}$. Namely, there are $V_{ \{ 00, 11\} } $ and $V_{ \{ 10, 01\} } $ that transform the Kraus basis independently. 

\begin{equation}
V_{ \{ 00, 11\} } = e^{-\frac{\theta(\bold{s})}{2}}\frac{ 2\sqrt{2} e^{ \frac{i}{2} (J'_z - J_z) }}{-1 - \cos(J'_x-J'_y) +e^{2 i J'_z}(-1 + \cos(J'_x + J'_y)) }
    \begin{pmatrix}
 V_{00} & V_{03}  \\ 
  V_{30} & V_{33} \\
\end{pmatrix}
\end{equation}
\begin{align*} 
V_{00} &= -e^{-\frac{s_{00}}{2}}( \cos(\frac{J'_x - J'_y}{2})\cos(\frac{J_x - J_y}{2}) + e^{i(J_z + J'_z)}\sin(\frac{J'_x + J'_y}{2})\sin(\frac{J_x + J_y}{2})  )\\
V_{03} &= ie^{-\frac{s_{00}}{2}}( -e^{iJ'_z}\sin(\frac{J'_x + J'_y}{2})\cos(\frac{J_x - J_y}{2}) + e^{iJ_z}\cos(\frac{J'_x - J'_y}{2})\sin(\frac{J_x + J_y}{2})  )\\
V_{30} &= ie^{-\frac{s_{11}}{2}}( -e^{iJ'_z}\sin(\frac{J'_x + J'_y}{2})\cos(\frac{J_x - J_y}{2}) + e^{iJ_z}\cos(\frac{J'_x - J'_y}{2})\sin(\frac{J_x + J_y}{2})  )\\
V_{33} &=  -e^{-\frac{s_{11}}{2}}( \cos(\frac{J'_x - J'_y}{2})\cos(\frac{J_x - J_y}{2}) + e^{i(J_z + J'_z)}\sin(\frac{J'_x + J'_y}{2})\sin(\frac{J_x + J_y}{2})  )\\
\end{align*}

\begin{equation}
V_{ \{ 10 , 01 \} } = e^{-\frac{\theta(\bold{s})}{2}}\frac{ 2\sqrt{2} e^{ \frac{i}{2} (J'_z - J_z) }}{\beta(1 - \cos(J'_x-J'_y) +e^{2 i J'_z}(1 + \cos(J'_x + J'_y)) )}
    \begin{pmatrix}
 V_{11} & V_{21}  \\ 
  V_{12} & V_{22} \\
\end{pmatrix}
\label{eq:basis-transformation-12}
\end{equation}
\begin{align*} 
\beta &= + \sqrt{ \frac{m_{00} - m_{33} +\sqrt{ (m_{00} - m_{33})^2 + 4m_{03}m_{30} }}{2m_{03}}}\\
V_{11} &= e^{-\frac{s_{10}}{2}}( \beta^2 e^{i(J_z + J'_z)}\cos(\frac{J'_x +J'_y}{2})\cos(\frac{J_x + J_y}{2}) + \sin(\frac{J'_x - J'_y}{2})\sin(\frac{J_x - J_y}{2})  )\\
V_{12} &= ie^{-\frac{s_{10}}{2}}(\beta^2 e^{iJ_z}\cos(\frac{J_x + J_y}{2})\sin(\frac{J'_x - J'_y}{2}) - e^{iJ'_z}\cos(\frac{J'_x + J'_y}{2})\sin(\frac{J_x - J_y}{2})  )\\
V_{21} &= ie^{-\frac{s_{01}}{2}}(e^{iJ_z}\cos(\frac{J_x + J_y}{2})\sin(\frac{J'_x - J'_y}{2}) - \beta^2 e^{iJ'_z}\cos(\frac{J'_x + J'_y}{2})\sin(\frac{J_x - J_y}{2})  )\\
V_{22} &=  e^{-\frac{s_{01}}{2}}(  e^{i(J_z + J'_z)}\cos(\frac{J'_x +J'_y}{2})\cos(\frac{J_x + J_y}{2}) + \beta^2 \sin(\frac{J'_x - J'_y}{2})\sin(\frac{J_x - J_y}{2})  )\\
\end{align*}

\section{Details for the example in the main text}
\label{sec:example}
In this section, we give the details of the dynamical crossover example discussed in the main text. For $\cos(J_z ) = 0$ and $s_{00} = s_{11}$, the tilted channel has a non-stationary steady state, the leading left eigenmatrix is a function of $s_{10} - s_{01}$ of the following diagonal form
\begin{equation}
    l^{1/2}_{\bold{s}} = \begin{pmatrix}
  \alpha & 0  \\ 
  0 & 1
\end{pmatrix},
\end{equation}
where $\alpha = (\frac{e^{-s_{01}}\cos^{2}(\frac{J_x+J_y}{2}) + e^{-s_{10}}\sin^{2}(\frac{J_x-J_y}{2}) }{e^{-s_{10}}\cos^{2}(\frac{J_x+J_y}{2}) + e^{-s_{01}}\sin^{2}(\frac{J_x-J_y}{2})})^{1/4}$. The Doob transformed Kraus operators are 
\begin{equation}
    \tilde{K}_{00} = \frac{e^{-s_{00}/2}}{e^{\theta(\bold{s})/2} } e^{-\frac{i \pi}{4}} \frac{1}{\sqrt{2}}
    \begin{pmatrix}
   \cos(\frac{J_x - J_y}{2}) & 0  \\ 
   0 & \sin(\frac{J_x + J_y}{2})
\end{pmatrix}
\quad
    \tilde{K}_{11} = \frac{e^{-s_{00}/2}}{e^{\theta(\bold{s})/2} } e^{-\frac{i \pi}{4}} \frac{1}{\sqrt{2}}
\begin{pmatrix}
  \sin(\frac{J_x + J_y}{2}) & 0  \\ 
   0 &  \cos(\frac{J_x - J_y}{2})
\end{pmatrix}
\end{equation}
and following the main text, we define $\mu = (s_{10}-s_{01})/2 $ and  $\delta=\sin((J_x -J_y)/2)/\cos((J_x+J_y)/2)$ for a compact notation 
\begin{equation}
    \tilde{K}_{10} = A(\mu,\delta)
    \begin{pmatrix}
  0 &  \sqrt{\delta^2+e^{2\mu}}  \\ 
  -\delta \sqrt{1+\delta^2e^{2\mu}}   & 0
\end{pmatrix}
\quad 
\tilde{K}_{01} = A(\mu,\delta)
    \begin{pmatrix}
  0 &  - \delta e^{\mu} \sqrt{\delta^2+e^{2\mu}}    \\ 
   e^{\mu} \sqrt{1+\delta^2e^{2\mu}}  & 0
\end{pmatrix},
\end{equation}
where 
\begin{equation}
    A(\mu,\delta) = \frac{ e^{\frac{i \pi}{4}} }{\sqrt{2}e^{\theta(\bold{s})/2}} e^{-s_{10}/2} \cos(\frac{J_x + J_y}{2}) \big(  (1 + \delta^2 e^{2\mu})( \delta^2 + e^{2\mu}) \big)^{-\frac{1}{4}}
\end{equation}
and the leading eigenvalue 
\begin{align}
\begin{split}
    e^{\theta(\bold{s})} &= \sqrt{(e^{-s_{10}}\cos^{2}(\frac{J_x + J_y}{2}) + e^{-s_{01}}\sin^{2}(\frac{J_x-J_y}{2}))(e^{-s_{01}}\cos^{2}(\frac{J_x + J_y}{2}) + e^{-s_{10}}\sin^{2}(\frac{J_x-J_y}{2}))} \\
    & \qquad \qquad + e^{-s_{00}}(\cos^{2}(\frac{J_x - J_y}{2}) + \sin^{2}(\frac{J_x+J_y}{2})) \\
    & = e^{-s_{10}} \cos^{2}(\frac{J_x + J_y}{2}) \sqrt{ (1 + \delta^2 e^{2\mu})( \delta^2 + e^{2\mu}) }  + e^{-s_{00}}(\cos^{2}(\frac{J_x - J_y}{2}) + \sin^{2}(\frac{J_x+J_y}{2}))
\end{split}
\end{align}

Furthermore, we focus on the case $s_{01} = -s_{10} $ and assuming $e^{-s_{00}}(\cos^{2}(\frac{J_x - J_y}{2}) + \sin^{2}(\frac{J_x+J_y}{2}))$ is order one, and show the asymptotic behavior of Doob transformed Kraus operators up to a complex phase in various limits
\begin{itemize}
    \item For $ e^{\mu}\ll \delta $, $e^{\theta(\bold{s})} \sim e^{-\mu} \delta $, $A(\mu,\delta) \sim \delta^{-1}$,  $ \tilde{K}_{10} \sim  \sigma_y $ and $ \tilde{K}_{01} \sim  e^{\mu} \delta^{-1}  \sigma_- $.
    \item For $ \delta \ll  e^{\mu} \ll  {\delta^{-1}}$, $e^{\theta(\bold{s})} \sim 1 $, $A(\mu,\delta) \sim e^{-\mu}$, $ \tilde{K}_{10} \sim \sigma^{+} $ and $ \tilde{K}_{01} \sim \sigma^{-} $.
    \item For $ e^{\mu}\gg {\delta^{-1}}$, $e^{\theta(\bold{s})} \sim e^{\mu} \delta $, $A(\mu,\delta) \sim  e^{-2\mu}\delta^{-1}$, $ \tilde{K}_{10} \sim e^{-\mu} \delta^{-1} \sigma_+ $ and $ \tilde{K}_{01} \sim  \sigma_y $,
\end{itemize}
where $\sim$ indicates proportionality with an unimportant constant of order one. 
\begin{figure}[t!]
    \centering
    \includegraphics[width=0.7\columnwidth]{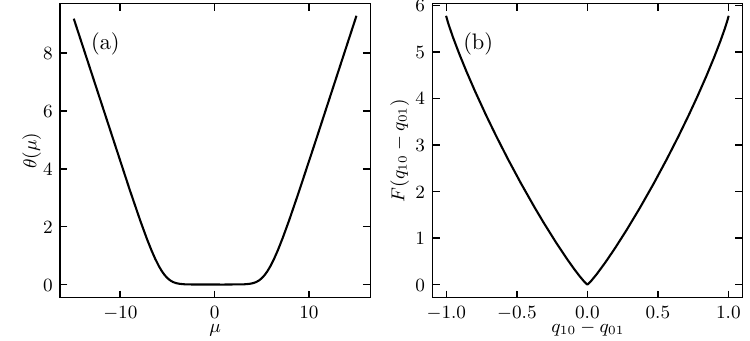}
    \caption{ a) The SCGF and b) rate function for the example discussed in the main text, evaluated for the same parameters as Fig. 3. }
    \label{fig:SM1}
\end{figure}

In Fig. \ref{fig:figure3}(a,b) of the main text, we have calculated the mean 
\begin{equation}
    \langle q_{10} - q_{01} \rangle = \frac{ \partial \theta(\bold{s}) }{\partial s_{01}} - \frac{ \partial \theta(\bold{s}) }{\partial s_{10}}
\end{equation}
and (normalised) variance 
\begin{equation}
     2t\, \text{Var}(q_{10} - q_{01}) =  \frac{ \partial ^2 \theta(\bold{s}) }{\partial s_{10} ^2} + \frac{ \partial^2 \theta(\bold{s}) }{\partial s_{01}^2} - 2\frac{ \partial ^2 \theta(\bold{s}) }{\partial s_{01} \partial s_{10} } 
\end{equation}
of the accumulated counting of $\{ 10 \}$ minus $ \{ 01 \}$. Here we plot the SCGF and rate function in Fig. \ref{fig:SM1}. The Kraus operators are different for the three asymptotic limits discussed above and are consistent with the mean accumulated counting. The SCGF shows a clear change in the gradient at the dynamical crossover point. 

In Fig. \ref{fig:figure3}(c), typical trajectories are shown for five different $\mu$. They are sampled by applying Eq.~(\ref{eq:update-prob}) at every time step. Starting from a Haar random initial state $|\psi_0 \rangle \langle \psi_0 |$ (which is the same for all five trajectories), 
the measurement outcome $ij$ occurs on step $t$ with probability
\be
\Tr[\tilde K_{ij} |\psi_t \rangle \langle \psi_t | \tilde K^{\dagger}_{ij}].
\ee
(These probabilities are normalised because $\sum_{ij} \tilde K_{ij}^\dag \tilde K_{ij} = \mathbf{1}$.)
If the measurement outcome is ${ij}$, the new normalised state at time $t+1$ is given by 
\begin{equation}
  |\psi_{t+1} \rangle \langle \psi_{t+1} | = 
 \frac{ \tilde K_{ij} |\psi_t \rangle \langle \psi_t | \tilde K^{\dagger}_{ij} }{\Tr[\tilde K_{ij} |\psi_t \rangle \langle \psi_t | \tilde K^{\dagger}_{ij}]} 
\end{equation}
similar to Eq. \eqref{eq:update-prob}.

In the large $\mu$ limit, the new circuit parameters are given by Eq. (\ref{eq:new_J}), which is a dual unitary circuit $(\pi/2, -\pi/2, \pi/2)$. The environmental unravelling defined by Eq. (\ref{eq:Computational-Kraus}) for $\{ 10 , 01  \}$ are represented by Kraus operators $(-1 + i)\sigma_y / 2 $ and $(1 - i)\sigma_y / 2$. The basis transformation to the Doob transformed is explicitly given by Eq. (\ref{eq:basis-transformation-12})
\begin{equation}
V_{ \{ 10, 01\} } = \frac{1}{\sqrt{2}}
    \begin{pmatrix}
 1 & 1 \\ 
  -1 & 1 \\
\end{pmatrix},
\end{equation}
and the Doob transformed Kraus operators are $\tilde{K}_{10} = 0$ and $\tilde{K}_{01} =(1 - i)\sigma_y / \sqrt{2}$. 

\section{General unitaries}
\label{sec:general_unitary}

The previous sections considered circuits where $U = U_{\text{XYZ}}$, i.e., $u_1,u_2,u_3,u_4 = \mathbf{1}$ in Eq. \eqref{eq:general_unitary} and the light-cone channel is given by Eq. (\ref{eq:Heisenberg-channel}). In this section, we provide some comments on the more general cases where the local unitaries $u_i=u(\theta_i,\phi_i,\psi_i)$, following Eq. (\ref{eq:smallu-parametrisation}). 

First, we note that only two local unitaries in Eq. (\ref{eq:general_unitary}) appear in the general light-cone channel Eq. (\ref{eq:general_U_channel}), so $u_2$ and $u_3$ do not affect the dynamics at the level of density matrices. However, $u_2$ and $u_3$ can affect the Kraus operators defined by Eq. (\ref{eq:Computational-Kraus}) so they change the statistics of the environmental measurements. This is equivalent to a change of measurement basis, just like the change of basis in Eq. (\ref{eq:Kraus-change-basis}).

On the other hand, the local unitaries $u_1$ and $u_4$ change the light-cone channel. A simple case is when $u_1 = u_4^{\dagger} $, and the local unitaries act as a similarity transformation on the XYZ unitaries [because $u_{1}u_{4} = \mathbf{1}$].  Then the spectrum of 
${\cal M}_{U,\bold{s}} = (u^{\dagger}_1 \otimes u_1^{T}) \mathcal{M}_{U_{\text{XYZ}},\bold{s}} ( u_1 \otimes u_1^{*} )$
matches that of $ \mathcal{M}_{U_{\text{XYZ}},\bold{s}}$, while the leading right eigenmatrix of $\mathcal{M}_{U,\bold{s}}$ is $u^{\dagger}_1 r_{\bold{s}} u_1 $, where $r_{\bold{s}}$ is the corresponding eigenmatrix of $\mathcal{M}_{U_{\text{XYZ}},\bold{s}}$.
Similarly the leading left eigenmatrix of $\mathcal{M}_{U,\bold{s}}$ is
\begin{equation}
l_{u} = u^{\dagger}_1 l_{\bold{s}} u_1
\end{equation}
Using Eq. (\ref{eq:Doob-Kraus}), one sees that the Doob channel becomes $\tilde{\cal M}_{U,\bold{s}}=e^{-\theta(\bold{s})}(l^{1/2}_{u} \otimes l^{1/2}_{u} ) \mathcal{M}_{U,\bold{s}} ( l^{-1/2}_{u} \otimes l^{-1/2}_{u} )$.

More generally, note that $(u^\dag \otimes u^T)\mathbf{1}=\mathbf{1}$ for any unitary $u$.  Hence if $l_{\bold{s}}=\mathbf{1}$ for a tilted XYZ channel $M_{U_{\text{XYZ}},\bold{s}}$ [that is, $ M_{U_{\text{XYZ}},\bold{s}}^\dag(\mathbf{1})=e^{\theta(\bold{s})}\mathbf{1}$] then one always has
\begin{equation}
{\cal M}_{U,\bold{s}}^\dag(\mathbf{1}) =  (u_1^\dag \otimes u_1^{T}) \mathcal{M}^\dag_{U_{\text{XYZ}},\bold{s}} ( u_4^\dag \otimes u_4^T )\mathbf{1} = e^{\theta(\bold{s})}\mathbf{1}
\label{equ:unital-u1-u4}
\end{equation}
That is, the leading left eigenmatrix of the tilted channel is the identity, whatever one takes for $u_1,u_4$.  
As discussed in the main text, the Doob transform tends to be simple in these cases.  This allows (for example) case A of the main text to be generalised to non-XYZ gates with arbitrary $u_1,u_4$.  The corresponding Doob-transformed channel lacks the symmetries $[(\sigma_i \otimes \sigma_i),{\cal M}_{U,\bold{s}}]=0$ but a $\tilde{U}$ can be found that reproduces its behavior, albeit not of XYZ form.

A similar (but weaker) result holds for all XYZ channels, because the leading eigenmatrix $l_{\bold{s}}$ of ${\cal M}_{U_{\text{XYZ}},\bold{s}}$ is always diagonal, see Eq. \eqref{equ:l-matrix}.  For $u_i=u(\theta_i=0,\phi_i,\psi_i)$ in the parameterisation of Eq. \eqref{eq:smallu-parametrisation} then $u_i$ is diagonal [and independent of $\psi_i$]. Then one has $(u_i^\dag \otimes u_i^T)l_{\bold{s}}=l_{\bold{s}}$ (diagonal matrices commute).  Repeating the argument of \eqref{equ:unital-u1-u4}, this implies that the leading left eigenmatrix of ${\cal M}_{U,\bold{s}}$ is unchanged by adding the (diagonal) local unitaries.
In this case the Doob-transformed channel commutes with $\sigma_z \otimes \sigma_z$ but not with $\sigma_x \otimes \sigma_x$.  This allows cases B and C of the main text to be generalised to suitable non-XYZ gates $U$, such that a suitable $\tilde{U}$ can be found (again with non-XYZ form).

Given these results, we end with one further comment. We have shown in Appendix \ref{sec:Doob_channel} that there are only 4 cases where Eq.~(\ref{eq:Doob-Heisenberg}) holds, assuming that $\tilde{\mathcal{M}}_{\bold{s}}$ is the Doob channel for a XYZ unitary and $\mathcal{M}(J_x', J_y' , J_z')$ is the light-cone channel for another XYZ unitary.  We discussed here some additional cases where analogous relations hold for non-XYZ unitaries. However, the full extension of Eq. (\ref{eq:Doob-Heisenberg}) to generic unitaries remains open: when can $\tilde{\mathcal{M}}_{U,\bold{s}} = \mathcal{M}_{\tilde{U}}$ be solved for $\tilde{U}$?  An obvious constraint is that $\mathcal{M}_{\tilde{U}}$ has to be unital, but this still leaves many possibilities for the unitaries and deformation parameters. For example, even if $U=U_{\rm XYZ}$, there might be solutions with $\tilde{U}\neq \tilde{U}_{\rm XYZ}$, for appropriate deformation parameters $\bold{s}$. Classifying the existence of solutions for $\tilde{U}$ requires a more complicated theory and remains an interesting direction for future work.

\end{appendix}
\twocolumngrid

\bibliography{bibliography-05082024.bib,extra}

\end{document}